\documentclass[twocolumn]{aastex701}
\usepackage{graphicx}
\usepackage{xspace}
\usepackage{xcolor}
\usepackage{amsmath}
\usepackage{mathrsfs}

\newcommand{\kms}{km\,s\ensuremath{^{-1}}\xspace}%
\newcommand{\masyr}{mas\,yr\ensuremath{^{-1}}\xspace}%
\newcommand{\mbh}{\ensuremath{M_{\bullet}}\xspace}
\newcommand{\vlos}{\ensuremath{V_\textrm{LOS}}\xspace}
\newcommand{\SgrA}{Sgr~A$^\star$\xspace}
\newcommand{\Agama}{\textsc{Agama}\xspace}
\newcommand{\F}{$\mathbb F$\xspace}
\newcommand{\V}{$\mathbb V$\xspace}

\begin{document}

\title{\mbox{Distribution function-based modelling of discrete kinematic datasets}, \mbox{in application to the Milky Way nuclear star cluster}}

\author[orcid=0000-0002-5038-9267]{Eugene Vasiliev}
\affiliation{University of Surrey, Guildford GU2 7XH, UK}
\email[show]{eugvas@protonmail.com}

\author[orcid=0000-0002-0160-7221]{Anja Feldmeier-Krause}
\affiliation{University of Vienna, T\"urkenschanzstra\ss e 17, 1180 Vienna, Austria}
\email{anja.krause@univie.ac.at}

\author[orcid=0000-0001-6113-6241]{Mattia C. Sormani}
\affiliation{Como Lake Centre for AstroPhysics (CLAP), DiSAT, Universit{\`a} dell'Insubria, Via Valleggio 11, 22100 Como, Italy}
\email{mattiacarlo.sormani@uninsubria.it}

\date{Received 2026 February 14; accepted 2026 March 27; published 2026 April 27}

\begin{abstract}
We present a method for constructing dynamical models of stellar systems described by distribution functions and constrained by discrete-kinematic data. 
We implement various improvements compared to earlier applications of this approach, demonstrating with several examples that it can deliver meaningful constraints on the mass distribution even in situations where the density profile of tracers and the selection function of the kinematic catalogue are unknown.
We then apply this method to the Milky Way nuclear star cluster, using kinematic data (line-of-sight velocities and proper motions) for a few thousand stars within 10 pc from the central black hole, accounting for the contributions of the nuclear stellar disc and the Galactic bar. We measure the mass of the black hole to be $4\times 10^6\,M_\odot$ with a 10\% uncertainty, which agrees with the more precise value obtained by the GRAVITY instrument. The inferred stellar mass profile depends on the choice of kinematic data, but the total mass within 10~pc is well constrained in all models to be (2.0--2.3)$\times 10^7\,M_\odot$. We make our models publicly available as part of the \Agama software framework for galactic dynamics.
\end{abstract}


\section{Introduction}

The centre of our Galaxy contains two exciting objects: the supermassive black hole (SMBH) \SgrA of mass $\mbh \approx 4.3\times 10^6\,M_\odot$ \citep{Gravity2022} and the surrounding nuclear star cluster (NSC). The NSC, in turn, is embedded in the nuclear stellar disc (NSD), which extends up to a couple hundred pc (see, e.g., \citealt{Schultheis2025} for a review), and at even larger distances (a few kpc), by the Galactic bar, then the Galactic stellar disc and halo, and finally the rest of the known Universe.

NSCs frequently coexist with SMBHs in galaxies of Milky Way mass or lower (see section 8 in \citealt{Neumayer2020}), but only in our Galaxy we have the luxury of being able to measure the SMBH mass from orbital fits to multi-epoch observations of stars in close proximity to it \citep[e.g.,][]{Ghez2008,Genzel2010}. In other stellar systems, the mass of the central SMBH is usually inferred from a single kinematic snapshot with integrated-light spectroscopic data of unresolved stellar populations. In rare cases (globular clusters or satellite galaxies of the Milky Way), one can work with line-of-sight velocity (\vlos) or proper motion (PM) measurements of resolved stars \citep[e.g.,][]{denBrok2014}, but even then the dynamical inference must rely on a single kinematic snapshot, which almost always requires an assumption of equilibrium to make progress in constraining the gravitational potential (mass distribution) of the system. 

In the Milky Way, individual stars tightly bound to \SgrA have been monitored over more than two decades, and some stars (notably, S2) have completed more than one orbit during this time, enabling very precise measurements of not only \mbh (at a sub-percent level), but even higher-order corrections, e.g., due to general relativity effects \citep{Gravity2018,Gravity2020}. By contrast, methods relying on the equilibrium modelling of the NSC stars at larger distances usually produce much larger uncertainties on the SMBH mass, and in many studies the inferred \mbh has been biased low by up to 50\% (see the introduction section in \citealt{Magorrian2019} for an overview of prior work, and figure~1 in \citealt{Eugene2026} for a more jocular literature analysis). Despite this mixed success in the galaxy we know best, values of SMBH masses in external galaxies determined from stellar kinematics often come with fairly small error bars (sometimes $\lesssim 10\%$), even though the quality of observational data is clearly inferior to the Milky Way. Except a few megamaser galaxies, such as NGC 4258 \citep{Miyoshi1995}, there are no other systems in which the SMBH mass can be measured with comparable accuracy to \SgrA to provide the ``true'' answer for benchmarking stellar-dynamical or other mass measurement methods.

The landscape of stellar-dynamical modelling methods is discussed in more detail in Section~\ref{sec:modelling_methods}, and in this work, we use the approach based on distribution functions (DFs) with specific analytic forms and tunable parameters (Sections~\ref{sec:df} and \ref{sec:scm}). Our models have separate DFs for the innermost component (NSC) and the surrounding NSD, but we fix the parameters of the NSD to the ones found in a recent paper \citep{Sormani2022a}. The latter study used kinematic data on larger scales (10--300~pc) and thus aimed at determining the parameters of the NSD while treating the NSC as a fixed mass component. While the modelling method used in that paper is largely the same, in the present work we have implemented significant upgrades that deserve a detailed description, which is given in Section~\ref{sec:model_fitting}. In particular, we consider a scenario in which the spatial coverage of kinematic data is non-uniform and not representative of the underlying stellar distribution, and the spatial density profile of kinematic tracers is unavailable. We demonstrate with several tests on mock data that one can derive meaningful constraints on the gravitational potential even in this case, which has received little attention in the previous literature. Our modelling workflow relies on the \Agama framework for galactic dynamics \citep{Vasiliev2019a}, and various technical details mentioned throughout the paper, such as class and parameter names, refer to the \texttt{Python} interface for \Agama.

The observational data used to constrain our models consist of a few thousand stars within 10~pc from \SgrA described in Section~\ref{sec:data}. This is a subset of the sample used in \cite{Feldmeier2025b}, but that work employed a different approach (axisymmetric Jeans models). Given that the data are shared between this and our work, we largely focus on the DF-based modelling approach, which has not been used for the NSC previously. As explained above, the NSC is not only interesting by itself, but also serves as a testbed for SMBH mass measurements from stellar kinematics.

This study thus has two main objectives: (1) introduce a practical method for construction of DF-based equilibrium dynamical models constrained only by discrete-kinematic data (i.e., line-of-sight velocities and PMs of individual stars, but not their density profile), and (2) apply it to the Milky Way NSC, test the accuracy of recovery of the SMBH mass, and provide the first DF-based model of this system. Sections~\ref{sec:model_construction} and \ref{sec:model_fitting} fulfill the first goal, and sections~\ref{sec:data} and \ref{sec:results} apply the method to the NSC. Section~\ref{sec:summary} summarises our findings.

\section{Construction of dynamical models}  \label{sec:model_construction}

\subsection{Overview of methods}  \label{sec:modelling_methods}

Several approaches have been used for modelling the Milky Way NSC and NSD in the past. Most studies relied on the Jeans equations, either spherical \citep[e.g.,][]{Schoedel2009, Do2013b, Chatzopoulos2015a, Fritz2016} or axisymmetric \citep{Feldmeier2014,Chatzopoulos2015a, Sormani2020, Feldmeier2025b}. These are easy to understand and fast to run, but have a few fundamental limitations. First, Jeans equations typically only use the first two moments of the velocity distribution; although there are implementations that can incorporate higher moments (e.g., \citealt{Merrifield1990, Magorrian1995, Richardson2013, Read2017, BanaresHernandez2026}), they have not been applied to the NSC. Second, a solution to the Jeans equations is not guaranteed to correspond to a physically valid (nonnegative) DF. This is rarely a concern for most stellar systems, but when the potential is dominated by a central point mass, the constraints imposed by the so-called ``cusp slope--velocity anisotropy theorem'' \citep{An2006} exclude a considerable part of the parameter space as unphysical. This may have contributed to an overestimated \mbh in anisotropic Jeans models of \citet{Do2013b}. The third limitation, namely, the well-known mass--anisotropy degeneracy in Jeans equations, is not so severe when one can measure more than one velocity component, as is the case for the NSC. Jeans models can be used with both binned and discrete-kinematic data, and rely on the density profile of tracers (typically determined independently of the kinematic sample) to solve the equations for the velocity moments.

At the other end of the spectrum of dynamical modelling methods (in terms of complexity and computational cost) is the \citet{Schwarzschild1979} orbit-superposition approach, in which the DF is modelled as a weighted sum of a large number of orbits (essentially, Dirac's $\delta$-functions in the space of integrals of motion). A triaxial orbit-based model of the NSC has been constructed by \citet{Feldmeier2017b}. This class of models is not well suited for dealing with discrete-kinematic data, necessitating their binning (apart from the proof-of-concept study of \citealt{Chaname2008}, there are no examples of ``discrete Schwarzschild models'' in the literature).

An approach that can be viewed as a crossover between orbit-based and DF-based methods is that of \citet{Magorrian2019}. In this case, the DF is represented by a weighted sum of $\sqcap$-shaped building blocks in the $E,L^2$ space, and the model fitting could be done using individual velocity measurements and evaluating the likelihood of each star against the DF of each building block. However, due to peculiarities of kinematic catalogues, \citet{Magorrian2019} had to seriously engage with the selection function, as discussed further down in Section~\ref{sec:fitting_general}. In his approach, the DF of stars did not contribute to the total gravitational potential (i.e., the model was not self-consistent in the sense of Section~\ref{sec:scm}); rather, the potential was created by the SMBH plus an extended mass profile, whose parameters were optimised during the fit.

In DF-based methods, one usually deals with smooth analytic functions rather than discrete blocks with finite support. As the arguments of the DF, one can use either the classical integrals of motion ($E$, $L$ or $L_z$) or actions (radial $J_r$, vertical $J_z$ and azimuthal $J_\phi$, see Section~3.5.3 in \citealt{BinneyTremaine} for definition). It is the latter variant that we will consider in this project, as action-based DFs are more general and can be used in both spherical and axisymmetric systems equally well. Although in some cases the contribution of kinematic tracers to the potential can be neglected (e.g., in dwarf spheroidal galaxies dominated by dark matter), in general one needs to determine the potential self-consistently from the DF, using the iterative procedure outlined in section~\ref{sec:scm}, possibly including external sources of potential, such as the SMBH. Spherical DF models have been used for constraining the dark matter profile in dwarf galaxies \citep{Pascale2018, Pascale2019, Pascale2025, ArroyoPolonio2025} and the mass of the central BH in the globular cluster NGC~104 \citep{DellaCroce2024} and Leo~I dwarf galaxy \citep{Pascale2024}. Non-spherical models of this kind have been constructed for the Milky Way disc \citep{Piffl2014, Binney2015, Cole2017, Binney2023, Binney2024} and halo tracers \citep{Williams2015b, Vasiliev2019b, Hattori2021, Wang2022}, some external galaxies \citep{GalanDeAnta2023}, and most relevant for our study, for the Milky Way NSD \citep{Sormani2022a}. In the latter work, the central mass concentration representing the NSC and the SMBH was modelled as a static component rather than described by a DF, and in the present study, we lift this limitation and endow the NSC with its own DF, whose parameters are optimised to match the observed kinematics of the central region.

\subsection{Distribution function-based models}  \label{sec:df}

In this work, we use separate DFs for the NSC and NSD. These are expressed in terms of actions, and use the St\"ackel fudge method \citep{Binney2012a} to convert from phase-space coordinates to actions.

The NSC is described by the \texttt{DoublePowerLaw} DF, which is similar to those introduced by \cite{Posti2015}, \cite{Williams2015a}, \cite{Pascale2018}, \cite{Binney2026}:
\begin{eqnarray}  \label{eq:df_nsc}
f(\boldsymbol J) &\propto&
\left[ 1 + \left(\frac{J_0}{h(\boldsymbol J)}\right)^\eta \right]^{\Gamma/\eta}
\left[ 1 + \left(\frac{g(\boldsymbol J)}{J_0}\right)^\eta \right]^{-\mathrm{B}/\eta} \\
&\times& \exp\left[ -\left(\frac{g(\boldsymbol J)}{J_0}\right)^2 \right]
\left(1 + \varkappa \tanh\frac{J_\phi}{J_{\phi,0}} \right), \nonumber\\
g(\boldsymbol J) &\equiv& g_r J_r +\, g_z J_z +\, (3-g_r-g_z) |J_\phi|, \nonumber\\
h(\boldsymbol J) &\equiv& h_r J_r + h_z J_z + (3-h_r-h_z) |J_\phi|, \nonumber
\end{eqnarray}

It is well suited for modelling spheroidal components, and has the following free parameters, listed in Table~\ref{tab:fiducial_model} along with their fiducial values and allowed range. The total mass $M_\text{NSC}$ is log-scaled in the fitting process, as are other dimensional parameters. The characteristic (scale) action $J_0$ defines the transition between the inner and outer asymptotic regimes. The power-law slope of the DF in action space in the inner and outer parts of the model is defined by $\Gamma$ and B, respectively. These parameters directly control the slope of the density profile generated by this DF, but the correspondence between the density slope and the DF slope is non-trivial and mediated by the potential. In the asymptotic regime of the Keplerian potential (both deep inside the SMBH sphere of influence or in the outskirts of the model), the density slope is d\,$\ln\rho$/d\,$\ln r = -(3+\Gamma)/2$ and likewise for B. For instance, $\Gamma=0$ corresponds to a density cusp $\rho\propto r^{-3/2}$, and negative values of $\Gamma$ produce shallower cusps; however, when $\Gamma<-1$, the cusp does not get shallower than $r^{-1}$. The steepness of the transition between the two asymptotic regimes is controlled by $\eta$, with larger values implying a sharper transition. Note that when other mass components in addition to the NSC itself are contributing to the potential, the relation between the DF parameters and the density profile that it generates becomes less straightforward.

The coefficients $h_{r,z}$ control the velocity anisotropy in the radial and vertical directions in the inner part of the model, and $g_{r,z}$ do the same in the outer parts. The relation between these coefficients and the ratio of velocity dispersions or the axis ratio $q\equiv z/x$ of the density profile is again mediated by other parameters in a complicated way. In the outer parts, larger $g_z$ produce more oblate models, and larger $g_r$ make the model more tangentially anisotropic. In the inner parts the trends are the same for $h_z$ and $h_r$ when the inner slope $\Gamma>0$, but are reversed when $\Gamma<0$; and when $\Gamma=0$, these coefficients cease to have any influence, which can be seen as a drawback of this family of DFs. Finally, the DF can describe a rotating system by setting the amount of asymmetry between prograde and retrograde orbits ($\varkappa$) and the characteristic angular momentum $J_{\phi,0}$, which determines how steeply the DF approaches the asymptotic amount of rotation: when $|\boldsymbol J_\phi|\ll J_{\phi,0}$, the asymmetry is very low, while in the opposite limit the ratio of prograde to retrograde stars approaches $(1+\varkappa)/(1-\varkappa)$. In addition, we impose an exponential cutoff of the DF at large actions by setting $J_\text{cutoff}=10^4$~pc\,\kms, gently truncating the model at $r\gtrsim100$ pc (well beyond the extent of our kinematic dataset).

\begin{table}
\caption{Parameters of the model (all except \mbh describe the NSC DF), their fiducial values, and the allowed range in the fitting process. Dimensional quantities are given in units of $10^6\,M_\odot$ for mass and 1 pc\,\kms for actions.
}  \label{tab:fiducial_model}
\hspace{-14mm} 
\begin{tabular}{llrc}
\hline
parameter & role & \makebox[0cm][r]{value} & range\\
\hline
$\log_{10} M_\text{NSC}$ & NSC mass (log-scaled)    & 1.78 & $1.3 \div 2.5$ \\
$\log_{10} J_0$ & characteristic action (log)\!\!   & 2.38 &   $1 \div 4$   \\
$\Gamma$        & inner power-law index             &$-1.0$&  $-2 \div 0.5$ \\
B               & outer power-law index             &  4.0 & $3.1 \div 8.0$ \\
$\eta$          & transition steepness              &  0.7 & $0.4 \div 3.0$ \\
$h_r$           & radial anisotropy (inner)         &  1.6 & $0.1 \div 2.8$ \\
$h_z$           & flattening (inner)                &  0.4 & $0.1 \div 2.8$ \\
$g_r$           & radial anisotropy (outer)         & 1.35 & $0.1 \div 2.8$ \\
$g_z$           & flattening (outer)                & 1.32 & $0.1 \div 2.8$ \\
$\varkappa$     & amount of rotation                & 0.95 &   $0 \div 1$   \\
$\log_{10} J_{\phi,0}$ & gradient of $v_\phi$ (log) &  2.2 &   $1 \div 4$   \\
\mbh                   & SMBH mass                  &  4.3 &   $2 \div 6$   \\
\hline
\end{tabular}
\end{table}

The NSD is described by the \texttt{QuasiIsothermal} DF, which is suited for disky components (e.g., \citealt{Binney2010}, \citealt{Binney2011}). Its parameters are fixed to those determined by \citet{Sormani2022a} by fitting the model to larger-scale kinematic data: total mass $M_\text{NSD}=970\times 10^6\,M_\odot$, disk scale length $R_\text{disk}=75$\,pc, scale height $h_\text{disk}=25$\,pc, central velocity dispersion $\sigma_{r,0}=75$\,\kms, the scale radius of the velocity dispersion $R_{\sigma_r}=10^3$\,pc.

\subsection{Construction of self-consistent models}  \label{sec:scm}

The DF $f$ is specified in terms of the integrals of motion in a given gravitational potential $\Phi$ -- in our case, actions $\boldsymbol J(\boldsymbol x, \boldsymbol v;\, \Phi)$. The density $\rho(\boldsymbol x)=\iiint \text{d}^3\boldsymbol v\; f\big(\boldsymbol J(
\boldsymbol x, \boldsymbol v)\big)$ is generated by the DF, and $\rho$ is  also connected to the gravitational potential via the Poisson equation  $\nabla^2\Phi=4\pi G\rho$. In a gravitationally self-consistent case, $\rho$, $\Phi$ and $f$ depend on each other (the latter through the dependence of actions on the potential). Such models can be constructed in an iterative approach \citep[e.g.,][]{Binney2014, Piffl2015, Sanders2015b, Binney2023}, which is implemented in the \Agama \texttt{SelfConsistentModel} class. The functional form and the parameters of the DF are fixed, while the density and the corresponding potential are updated in the course of iterations. We refer to  Section~5.2 in \cite{Vasiliev2019a} for a full description of the method, and outline specific details of our setup below.

In our case, the model consists of three components: the static Newtonian potential of the SMBH and two DF-based components (NSC and NSD). Given a potential, each of these models produces a certain axisymmetric density profile $\rho(R,\theta)$, represented by a \texttt{DensitySphericalHarmonic} expansion of order $\ell_\text{max}$ in $\theta$, with coefficients depending on the spherical radius $r$. We use $\ell_\text{max}=6$ for the NSC and $\ell_\text{max}=12$ for the NSD, and a logarithmically-spaced grid with 25 points for each component (between 0.01 and 1000 pc for the NSC, and between 0.1 and 1000 pc for the NSD). The potential is computed using the \texttt{Multipole} expansion, which is well suited for cuspy density profiles that are not too flattened. We note that the multipole expansion is not very accurate for the NSD at large radii, and an alternative \texttt{CylSpline} potential should in general be used for disky components (as in \citealt{Sormani2022a}). Ideally, one could use both potential expansions (each for its own component), but this would double the cost of model construction for little extra benefit, since our kinematic data are confined to the central region, where \texttt{Multipole} is accurate enough.

The iterative process starts by initializing the potential to the same ``seed'' one used by \cite{Sormani2022a}, since we use their best-fit DF for the NSD component. An inherent deficiency of the \texttt{QuasiIsothermal} DF used for the NSD is that the final model (after the iterations) still incorporates some information from the ``seed'' potential used for initialisation, namely, the radial profiles of epicyclic frequencies in this ``seed'' potential are absorbed into the definition of the DF. The alternative \texttt{Exponential} DF used in \citet{Binney2023} is entirely potential-independent, but for the sake of compatibility with the previous NSD model, we keep the same DF. The ``seed'' potential is built from both NSD and NSC density profiles taken from \cite{Sormani2020} and \cite{Chatzopoulos2015a}, respectively, which are \textit{not} identical to the final density profile of the self-consistent model. The model DFs are then initialised from the provided parameters (and in the case of the NSD DF, the ``seed'' potential).

In each iteration, the density of both components is recomputed from their DFs in the current potential (which is the most expensive operation, since it involves the integration of the DF over the 3d velocity space, and each DF evaluation involves the conversion from the phase-space coordinates $\boldsymbol x,\boldsymbol v$ to actions $\boldsymbol J$). The potential is then recomputed from the updated density profiles. Since the NSD density is subdominant in the inner region covered by our kinematic data, we can save computational effort by keeping a fixed NSD density profile identical to the final model of \cite{Sormani2022a} for the first four iterations (this is not the ``seed'', which is only used for initialisation), and replacing it with the DF-based density only on the final iteration, when the DF-based NSC density has largely converged. We checked that increasing the number of iterations changes the model properties only marginally.

After we obtain the final density and potential profiles for the current choice of DF parameters, we evaluate the likelihood of the model, as described in the next section.

\section{Model fitting}  \label{sec:model_fitting}

Throughout this work, we denote the position and velocity in the \textit{intrinsic} coordinate system of the model by lowercase $\boldsymbol x, \boldsymbol v$, and in the \textit{observed} system by uppercase $\boldsymbol X, \boldsymbol V$. In the application to the NSC, we adopt the standard right-handed Galactocentric coordinate system for the model, shifted slightly to place the SMBH exactly at origin, with $xy$ being the equatorial plane and the Sun located at $x=-D_0\approx -8.2$~kpc \citep{Gravity2019}. Although there are indications that the major axis of the NSC is slightly tilted with respect to the Galactic plane \citep{Feldmeier2014,GallegoCano2020}, we have to ignore this misalignment, since both NSC and NSD must share the same coordinate system in an axisymmetric equilibrium dynamical model. The observed coordinate system is also centered on \SgrA, but is rotated such that $Z$ is the direction along the line of sight (hence $V_Z\equiv \vlos$), $Y$ points towards the Galactic north, and $X$ lies in the Galactic plane and increases with the longitude $l$; thus $x=Z, y=X, z=Y$. Given the small size of the NSC compared to the distance $D_0$, we convert PMs to velocities using a fixed value of $D_0$; thus $V_X\equiv V_\text{hor} \equiv 4.74\, D_0\, (\mu_l-\mu_l^{\text{SgrA}^\star})$ and $V_Y \equiv V_\text{ver} \equiv 4.74\, D_0\, (\mu_b-\mu_b^{\text{SgrA}^\star})$, where 4.74 is the conversion factor from \masyr to \kms. The PMs are measured relative to \SgrA, using the values for the latter quoted in \cite{Reid2020}.

\subsection{General considerations}  \label{sec:fitting_general}

One advantage of DF-based models is that they provide a direct way to compute the likelihood of any observation in the given model. Indeed, if all six phase-space coordinates of a star are known precisely, then $f\big(\boldsymbol J(\boldsymbol x, \boldsymbol v;\;\Phi)\big)$ is the probability of finding such a star in the model. Since the DF depends on actions $\boldsymbol J$ (or more generally, any integrals of motion), which in turn depend on the potential $\Phi$, one can expect that maximising the joint log-likelihood of the entire observed catalogue,
\begin{equation}
\ln \mathcal L \equiv \sum_{i=1}^{N_\text{stars}} \ln \mathcal L_i = \sum_{i} \ln f(\boldsymbol x_i, \boldsymbol v_i) ,
\end{equation}
will provide the best-fit parameters of both the DF and the potential (in a self-consistent model, the latter is determined by the former, but in general, we have other components in the potential, such as the SMBH). This is indeed true, but there are some complications.

First, when not all phase-space coordinates are known precisely (or at all), one needs to marginalise the DF over missing coordinates and/or measurement uncertainties. In our application, the distance to star $Z$ is unknown, some stars have no PM measurements, and all three velocity components come with uncertainties (we assume that errors follow the normal distribution $\mathcal N$). A \textit{projected DF} for such a star with coordinates $X,Y$ in the sky plane, measured line-of-sight velocity $V_Z$ and its uncertainty $\epsilon_{V_Z}$ would be
\begin{equation}  \label{eq:projected_df_cubature}
\!\!\!\!\!\!\begin{array}{c}
\displaystyle f_\text{proj}(X, Y, V_Z) = \int_{-\infty}^{\infty}\!\! dZ \int_{-\infty}^{\infty}\!\! dV_X \int_{-\infty}^{\infty}\!\! dV_Y \int_{-\infty}^{\infty}\!\! dV_Z'\\[3mm]
\displaystyle f(X,Y,Z,V_X,V_Y,V_Z')\; \mathcal N\big(V_Z'\,|\,V_Z,\epsilon_{V_Z}\big) .
\end{array}
\end{equation}
The above expression assumes that the line-of-sight thickness of the system is small compared to the distance $D_0$ to its centre (i.e., uses a Cartesian rather than spherical projection); in the general case, one would need to integrate along the distance $D\ge 0$ rather than $Z\in (-\infty..\infty)$ and include a weighting factor $D^2$ for the spatial density and another factor $D^2$ converting PMs to velocities $V_{X,Y}$.

\begin{figure*}
\caption{\noindent\normalsize Illustration of the fitting process in three scenarios, depending on the distribution of stars with kinematic measurements (black dots) and without (red). In panel A, the observed sample traces the entire population, i.e., the selection probability is uniform across the entire system. In panel B, only a \href{https://galileo-unbound.blog/2019/06/16/vladimir-arnolds-cat-map/}{certain region} is observed (the spatial selection function $S(\boldsymbol x)$ is zero outside the region, and needs not be uniform within it, but at least is known in advance). In panel C, stars are selected for kinematic observations based on some criterion that is difficult or impossible to formalise, making the selection function uncomputable.\\[1mm]
\includegraphics{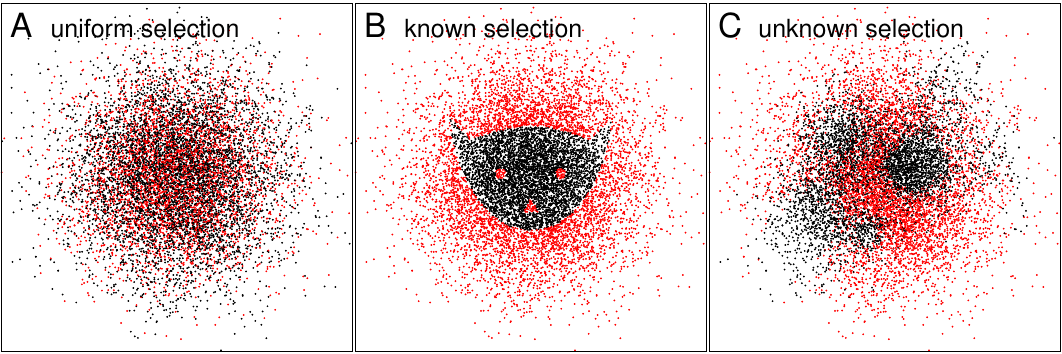}
\begin{tabular}{p{2.2in}p{2.2in}p{2.25in}}
\large$\displaystyle \ln\mathcal L = \sum_{i=1}^{N_\text{stars}}
\ln f(\boldsymbol x_i, \boldsymbol v_i)$. &
\large$\displaystyle \ln\mathcal L = \sum_{i=1}^{N_\text{stars}}
\ln \frac{f(\boldsymbol x_i, \boldsymbol v_i)}{\mathscr N}$, &
\large$\displaystyle \ln\mathcal L = \sum_{i=1}^{N_\text{stars}}
\ln \mathfrak{f}(\boldsymbol v_i \:|\: \boldsymbol x_i)$, \\
&
\normalsize $ \mathscr N \equiv 
\iiint \text{d}^3 \boldsymbol x \iiint \text{d}^3\boldsymbol v\; 
f(\boldsymbol x, \boldsymbol v)\; S(\boldsymbol x)$. &
\normalsize $\displaystyle \mathfrak{f}(\boldsymbol v \:|\: \boldsymbol x) \equiv 
\frac{f(\boldsymbol x, \boldsymbol v)}{\iiint \text{d}^3 \boldsymbol v\; f(\boldsymbol x, \boldsymbol v)} = \frac{f(\boldsymbol x, \boldsymbol v)}{\rho(\boldsymbol x)}$.
\end{tabular}\\[2mm]
Note that in the first two cases, $f(\boldsymbol x, \boldsymbol v)$ is the \textit{joint} probability of observing a star with given phase-space coordinates, whereas in the last case, $\mathfrak{f}(\boldsymbol v \:|\: \boldsymbol x)$ is the \textit{conditional} probability (i.e., the velocity distribution function at a given position). $\mathscr{N}$ is the normalisation factor, i.e., the integral of $f$ over the entire phase space, weighted by the selection function $S$ of the observed sample, which is usually purely spatial (independent of velocity). In the case of a complete or at least a uniform coverage ($S(\boldsymbol x)$=const), the second case is equivalent to the first one, but the third one is not: it entirely ignores the spatial distribution of stars, since one cannot assume that they faithfully trace the entire population. However, if this is actually the case, one can add another term in the likelihood function corresponding to the spatial probability, i.e., $\ln\mathcal L = \sum_{i=1}^{N_\text{stars}} \ln \rho(\boldsymbol x_i)$, and then the result becomes equivalent to the first case.
} \label{fig:selection_function}
\end{figure*}

The second, even more serious hurdle is that the observational sample is only a \textit{subset} of all stars described by the DF. 
The implications of this factor vary in complexity, as illustrated by three scenarios A, B \& C in Figure~\ref{fig:selection_function}. The general principles and some practical considerations of DF fitting in the presence of observational selection are discussed in \cite{McMillan2012,McMillan2013,Bovy2013,Ting2013,Trick2016}. Obviously, faint stars would be missing from the observational sample anyway, but if it were magnitude-complete and spatially uniform, this would not be a problem since the probability of a star being included in the observational catalogue (a.k.a.\ \textit{selection function}, SF) would be independent of phase-space coordinates (Scenario A, left panel). However, often observations only cover some fraction of the system, so that some portions of phase-space are more likely to be observed than others, in which case the SF $S(\boldsymbol x)$ is a function of sky-plane position and possibly distance. In this case (Scenario B, middle panel), one needs to adjust the likelihood computation to account for the fraction of the entire phase space that can appear in the catalogue (known as the \textit{normalisation factor} $\mathscr N$):
\begin{subequations}
\begin{eqnarray}
\mathcal L_i &=& f(\boldsymbol x_i, \boldsymbol v_i) \big/ \mathscr N, \\
\mathscr N &\equiv& \iiint \mathrm{d}^3\boldsymbol x \iiint \mathrm{d}^3\boldsymbol v\; f(\boldsymbol x, \boldsymbol v)\; S(\boldsymbol x) .
\end{eqnarray}
\end{subequations}
This approach was used, e.g., in \citet{Hattori2021} to compute the volume accessible to a magnitude-complete sample of RR Lyrae stars in the Galactic halo, excluding the low-latitude region and accounting for the scatter in the absolute magnitudes (i.e., for a soft truncation of the sample with distance). However, the SF of spectroscopic surveys is usually much more complicated \citep[e.g.,][]{Wojno2017,Imig2025}, and sometimes is practically uncomputable. Nevertheless, if it is independent of velocities (as is usually the case), one can instead work with \textit{conditional probability} $\mathfrak{f}(\boldsymbol v_i\, |\, \boldsymbol x_i)$ that the $i$-th star has a velocity $\boldsymbol v_i$ given its position $\boldsymbol x_i$ (Scenario C, right panel of Figure~\ref{fig:selection_function}). As per Bayes's rule, this conditional probability is given by the \textit{joint probability} in the six-dimensional phase space $f(\boldsymbol x, \boldsymbol v)$, divided by the marginalisation of the latter over velocities (which is just the spatial density of tracers $\rho$ at the given position):
\begin{equation}  \label{eq:conditional_likelihood}
\mathcal L_i = \mathfrak{f}\big(\boldsymbol v_i\, |\, \boldsymbol x_i\big) =
\frac{f(\boldsymbol x_i, \boldsymbol v_i)}{\iiint \mathrm{d}^3\boldsymbol v\; f (\boldsymbol x_i, \boldsymbol v)} =
\frac{f(\boldsymbol x_i, \boldsymbol v_i)}{\rho(\boldsymbol x_i)} .
\end{equation}

When the distance to stars is also unknown (as in our case), the above expressions must be marginalised over it (Section~\ref{sec:projected_df}). Here, we make another simplifying assumption that the SF does not depend on the distance. This is not strictly true for the NSC, because the extinction in this region is highly variable on small spatial scales \citep{NoguerasLara2021a}. As a result, stars separated in distance by only tens of pc can have substantially different probabilities of being observed. In practice, this leads to the asymmetry in the PM distributions in the horizontal direction (\citealt{Chatzopoulos2015b}; see also Figure~\ref{fig:velocity_distributions} in the Appendix). Section~\ref{sec:3d_extinction} further discusses the implications of this assumption and possible mitigation strategies.

Scenario C is particularly interesting, since it enables model fitting using only the kinematic information, but not the positions of stars. We note that in discrete-kinematic Jeans models, one computes the likelihood of stars against a Gaussian velocity distribution with mean and dispersion given by the Jeans equations, but the latter still require the density of tracers to be specified, even if it does not enter the likelihood function. On the other hand, it is possible to add another term in the likelihood function that accounts for the spatial density of tracers from a photometric catalogue (in turn, presumed to be complete or have a uniform SF, as in scenario A):
\begin{equation}
\ln \mathcal L^\text{(phot)} = \sum_{i=1}^{N_\text{stars}^\text{(phot)}} \ln \rho(\boldsymbol x_i),
\end{equation}
It is easy to see that if this sample coincides with the kinematic catalogue, the total log-likelihood reverts to scenario A, and in the more general case when the two samples are distinct, this extra term can still help to constrain the model.

Previous work on DF-based modelling used various approaches to dealing with SFs. For instance, \citet{ArroyoPolonio2025} estimated the spatially-varying SF by comparing the numbers of stars in the spectroscopic and photometric samples (the latter presumed to be complete down to some magnitude), then followed scenario B of a known SF.
In the application of DF models to the Galactic disc, \citet{Binney2023,Binney2024} sidestepped the need to work with SFs by computing the conditional probability of stellar velocities against velocity distribution functions (VDFs) at the given position provided by the model (scenario C), but they still added constraints on the density of tracers obtained from a different photometric sample. To our knowledge, the only attempt to fit dynamical models not explicitly constrained by any information about the density profile was the DF-based model of the NSD by \citet{Sormani2022a}.

\subsection{Toy examples}  \label{sec:toy_examples}

How far can one get with only kinematic information, anyway? Here we present two toy examples illustrating that one can meaningfully constrain the potential without using any information about the spatial distribution of tracers.

\subsubsection{1d vertical disc profile}  \label{sec:toy_1d}

As a first example, we revisit the classical problem of inferring the vertical mass distribution in a Galactic disc. Ignoring the radial gradients in the potential and the tracer distribution, the problem reduces to one dimension: measure $\Phi(z)$, or equivalently, the vertical density profile $\rho(z)$, using the $z$-velocity component of stellar tracers and possibly their $z$-coordinates. We do not assume any link between the tracer density profile $\rho_\star(z)$ and the mass density $\rho(z) \equiv (4\pi\,G)^{-1} \mathrm{d}^2\Phi/\mathrm{d}z^2$; in other words, the system is not gravitationally self-consistent, but is only assumed to be in a hydrostatic equilibrium.

The most well-known approach to this problem is the 1d vertical Jeans equation, which can be written as
\begin{equation}  \label{eq:jeans1d}
\frac{\mathrm{d}\Phi}{\mathrm{d}z} = -\frac{1}{\rho_\star} \frac{\mathrm{d}(\rho_\star\,\sigma^2)}{\mathrm{d}z},
\end{equation}
where $\sigma$ is the velocity dispersion of tracers. In this approach, one needs to know the vertical profiles of both $\rho_\star$ and $\sigma$ to measure $\Phi$.

An alternative approach, introduced by \cite{Kuijken1989}, is based on the Jeans theorem, representing the DF of tracers as some function $f(E)\equiv f\big(\Phi(z) + \frac{1}{2}v^2\big)$. If this DF can be measured locally (at $z=0$), where it is essentially the 1d VDF $f(v)|_{z=0}$, one can link the density it produces at some point $z$ to the potential difference between $z=0$ and that point, namely, $\rho_\star(z) = \int_{-\infty}^\infty f\big(\Phi(z) + \frac{1}{2}v^2\big)\, \mathrm{d}v$. By comparing this density profile with the actually measured one, the potential can be constrained without knowing anything about the kinematics of stars at $z>0$. In practice, it may be limited by the fact that the inference on the potential at high $z$ relies upon the behaviour of $f(v)$ in the tails (i.e., the energetic part of the population that can reach high values of $z$), which can be difficult to measure.

\begin{figure*}
\includegraphics{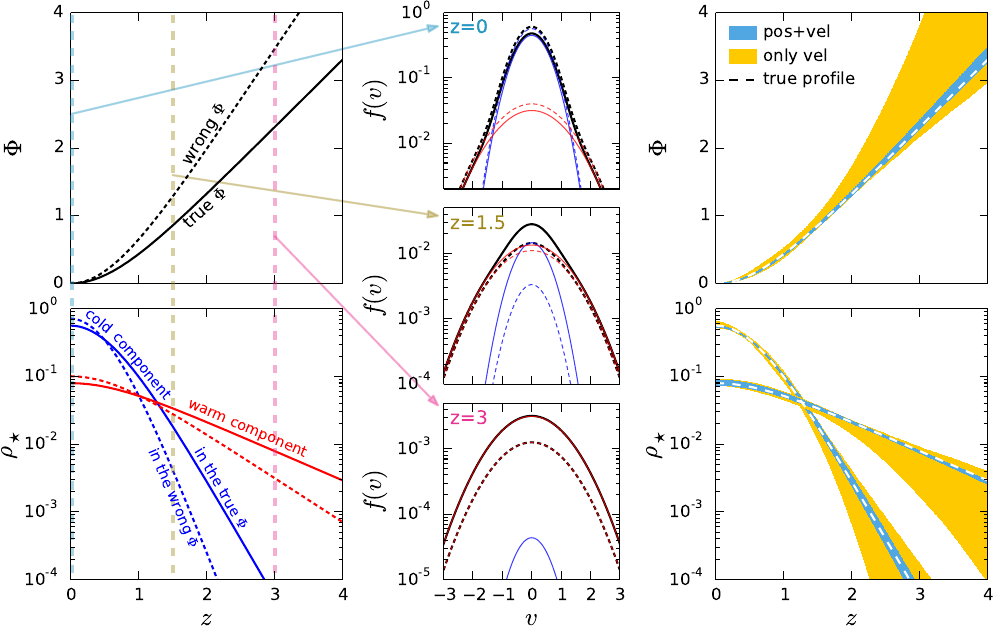}
\caption{Illustration of constraining the potential using the 1d toy model (Section~\ref{sec:toy_1d}) of a two-component tracer population with different velocity dispersions. Left and centre column demonstrate the basic idea: the density profiles $\rho_\star(z)$ of both components depend on the potential $\Phi(z)$ through Equation~\ref{eq:df1d_rho}, but this dependence is stronger for the colder component (blue) than for the warmer one (red). Centre column shows the velocity distribution functions (VDFs) at three values of $z$ (marked by vertical dashed lines in the left column), separately for the cold and warm components (blue and red Gaussians), as well as their sum; solid lines correspond to the true potential, and dashed -- to a different (wrong) potential. The amplitudes of the two Gaussians are the density values at the given $z$, so they change depending on the potential. At $z=0$ (top row) the potential is the same, and so are the VDFs (up to a small difference in the overall amplitude). At an intermediate value $z=1.5$ (middle row), the density of the cold component is much lower in the wrong potential than in the true one, so the shape of the VDF is sensitive to the choice of $\Phi$. At an even higher value $z=3$ (bottom row), the cold component makes negligible contribution in both potentials (the dashed blue line is not even visible on this plot), so the VDF is again nearly the same in both cases up to a difference in the overall amplitude. This suggests that one can constrain the potential from the shape of the VDF, as long as there is a mixture of populations with different temperatures.\protect\\
Right column shows the results of fitting the mass profile specified by Equation~\ref{eq:df1d_trialdensity} to the sample of $10^4$ tracers, using both their positions and velocities in the likelihood function (Equation~\ref{eq:df1d_jointposvel}, cyan contours) or only the velocities conditioned on the position (Equation~\ref{eq:df1d_onlyvel}, yellow contours). The shaded regions correspond to 16/84\% confidence intervals of the inferred potential (top) and density profiles of both components (bottom), while the dashed lines show the true profiles (same as solid lines in the left column). It demonstrates that one can get rather tight constraints in the region where the two kinematically distinct components overlap, but the uncertainties explode at $z\ge 2$ where only one (warmer) component is dominant.
}  \label{fig:toy_1d}
\end{figure*}

This method works with any DF, but it is instructive to consider a Maxwellian DF $f(E) \propto \exp\big(-E/\sigma^2\big)$ with a constant $\sigma$. In this case, $\rho(z) \propto \exp\big(-\Phi(z)/\sigma^2\big)$, and taking its logarithmic derivative by $z$, we recover the Jeans equation (\ref{eq:jeans1d}). Obviously, the velocity distribution remains Gaussian at any $z$ regardless of $\Phi$; what changes is only its normalisation (i.e. the tracer density $\rho_\star$), so without measuring it, one cannot constrain the potential at all.

Consider now a situation when the DF is a sum of two Maxwellian components with different velocity dispersions:
\begin{subequations}  \label{eq:df1d}
\begin{equation}  \label{eq:df1d_fE}
f(E) = A_1\,\exp\big(-E/\sigma_1^2\big) + A_2\,\exp\big(-E/\sigma_2^2\big).
\end{equation}
The corresponding density profile is $\rho_{\star}\equiv \rho_{\star,1} + \rho_{\star,2}$ with
\begin{equation}  \label{eq:df1d_rho}
\rho_{\star,c}(z) = \sqrt{2\pi} A_c\,\sigma_c\,\exp\big(-\Phi(z)/\sigma_c^2\big),\quad c=1,2,
\end{equation}
and the total mass of each component is
\begin{equation}  \label{eq:df1d_norm}
M_{\star,c} = \sqrt{2\pi} A_c\,\sigma_c\; 2\int_0^\infty \exp\big(-\Phi(z)/\sigma_c^2\big)\; \mathrm{d}z.
\end{equation}
\end{subequations}

Now the VDF at any point $z$ is a sum of two Gaussians, but with their relative amplitudes proportional to the density of the corresponding component $\rho_{\star,c}(z)$, which do depend on the potential. So one could expect that the potential can be constrained using the shape of the VDF, even if we do not know its normalisation (i.e. the total tracer density $\rho_\star$).
This is illustrated in the left and central column of Figure~\ref{fig:toy_1d}.

Here we test this scenario using the following setup: the true potential is
\begin{subequations}
\begin{equation}
\Phi^\text{true} = \ln\cosh(z/h^\text{true}), \quad h^\text{true}=1,
\end{equation}
and the corresponding mass density profile is
\begin{equation}
\rho^\text{true}(z) = \big[h^\text{true}\,\cosh(z/h^\text{true})\big]^{-2} / (4\pi\,G).
\end{equation}
For the fit, we use a mass density profile
\begin{equation}  \label{eq:df1d_trialdensity}
\rho(z) = \rho_0\, \Big[\cosh\big( [z/h]^\alpha \big)\Big]^{-\beta},
\end{equation}
which satisfies the natural requirement to be positive and monotonically decreasing with $z$, but is otherwise sufficiently flexible with four free parameters $\rho_0, h, \alpha, \beta$. The potential $\Phi(z) = 4\pi\,G \int_0^z \rho(z')\;(z-z')\;\mathrm{d}z'$ is computed by numerical integration. The positions and velocities for $N_\text{stars}=10^4$ tracers are drawn from a mixture of two Maxwellian DFs (\ref{eq:df1d}) with $M_{\star,1}=0.75$, $M_{\star,2}=0.25$, $\sigma_1=0.5$, $\sigma_2=1.0$, and in the fit, we assume the same functional form, but with $M_{\star,1}$, $\sigma_1$ and $\sigma_2$ being free parameters, and fixing $M_{\star,2} = 1 - M_{\star,1}$.
\end{subequations}

\begin{subequations}
Figure~\ref{fig:toy_1d}, right column, shows the fit results in two cases.
In the first run, we constrain the model by the joint likelihood of tracers in the position and velocity space:
\begin{equation}  \label{eq:df1d_jointposvel}
\ln\mathcal L = \sum\nolimits_{i=1}^{N_\text{stars}} \ln f\big( \Phi(z_i) + {\textstyle\frac{1}{2}} v_i^2 \big),
\end{equation}
where $f(E)$ is given by Equation~\ref{eq:df1d_fE}; in other words, use both the position and velocity data for each star.
In the second one, we use the conditional likelihood function for the velocity space only, as in Equation~\ref{eq:conditional_likelihood}, i.e., fit only the shape the VDF, but not its normalisation:
\begin{equation}  \label{eq:df1d_onlyvel}
\ln\mathcal L = \sum\nolimits_{i=1}^{N_\text{stars}} \ln \big[ f\big( \Phi(z_i) + {\textstyle\frac{1}{2}} v_i^2 \big) \;\big/\; \rho_\star(z_i) \big].
\end{equation}
\end{subequations}

In both runs, the DF parameters ($M_{\star,1}, \sigma_1, \sigma_2$) are well measured with uncertainties of only a few percent: this can be achieved without considering the potential at all, just by fitting a two-component Gaussian mixture model to the velocity distribution of all tracers regardless of their positions. Clearly, in the first case one can also constrain the potential and the tracer density profiles of both components fairly well. Interestingly, the \cite{Kuijken1989} approach produces very similar uncertainties, demonstrating that it does not matter much whether we know the velocity distributions at all $z$ or only at $z=0$, as long as the tracer density profile at all $z$ is known. However, even in the second case one can still obtain quite strong constraints on the potential up to $z\lesssim 2$ (in the region when both Maxwellian components contribute to the VDF), while at higher $z$ the uncertainties explode, since only the second (warmer) component survives and the VDF becomes nearly Gaussian. This confirms our expectations that a non-Gaussian VDF is sufficiently informative by itself (this was hinted at the end of section~4.4 in \citealt{Binney2024}). On the other hand, it constrains only the total potential, and if one is interested in separating the contribution of stars and dark matter, then one would likely need to also fit for the stellar density profile (as in the application of DF modelling to the Milky Way disc by \citealt{Binney2015,Binney2023,Binney2024}, who used the observed vertical density profile in addition to the kinematic catalogue).

It is worth pointing out similarities and differences between this example and the \cite{Walker2011} approach for constraining the potential of a spherical system using two tracer populations. In the latter approach, the inference relies on a simple virial mass estimator that provides the value of the enclosed mass at a ``sweet-spot'' radius comparable to the half-light radius of the tracer population, where the impact of the (unknown) velocity anisotropy of tracers is minimised. When there are multiple tracer populations with different half-light radii, one can constrain the potential at several points and thus estimate the slope of the mass density profile in this radial range. The idea of using multiple tracer populations to better constrain the potential can be extended to other dynamical modelling approaches, such as Jeans \citep{Zhu2016} and DF-based models \citep{ArroyoPolonio2025}, but in all cases, it is not a critically important ingredient. By contrast, in our toy example a single Maxwellian DF would not enable any inference on the potential; however, a single population with a more complicated DF could still produce VDFs that are somewhat sensitive to the potential.

\subsubsection{Spherical halo models}  \label{sec:toy_spherical}

\begin{figure*}
\includegraphics{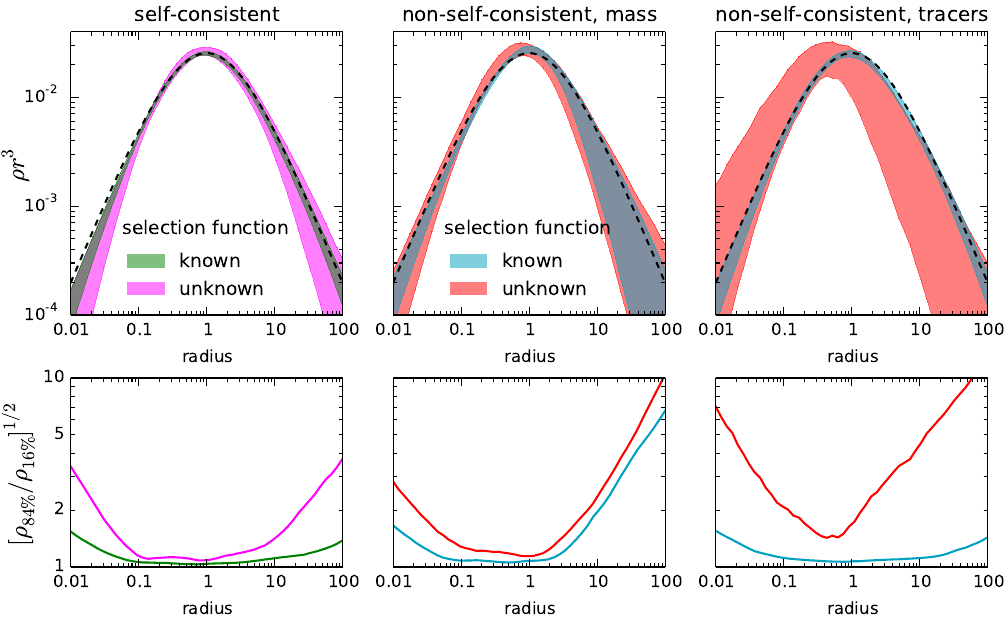}
\caption{Illustration of the dynamical modelling precision in different scenarios using the toy spherical model (Section~\ref{sec:toy_spherical}). Shown are constraints on the mass density (left and middle columns) and tracer density (right column), in the cases when the spatial selection function is known and uniform (scenario A in Figure~\ref{fig:selection_function}; green and cyan) or unknown (scenario C, which uses only the kinematic information but not the spatial distribution of tracers; magenta and red curves). In the self-consistent case (left column), tracer and mass densities are identical, while in the opposite case, they are shown separately (middle and right columns). The top row shows the true profile as a black dashed line, and the 16/84\% confidence intervals on $\rho(r)\,r^3$ as shaded regions, and bottom row shows the square root of the ratio of upper and lower bounds of these intervals (qualitatively, the relative uncertainty, with 1 corresponding to a precise constraint and $\gg1$ to a poor constraint). Naturally, the constraints are tighter at intermediate radii where the majority of tracers reside ($\rho r^3$ shows the mass of stars per logarithmic interval in radius). The radial range roughly encompasses all $10^3$ kinematic tracers used in the fit.
}  \label{fig:toy_spherical}
\end{figure*}

Next, we consider a toy model of a single-component spherical system with a double-power-law density profile \citep{Zhao1996}:
\begin{equation}
\rho(r) = \rho_0\; \left(\frac{r}{r_0}\right)^{-\gamma} \left(1 + \left[\frac{r}{r_0}\right]^\alpha\right)^{(\gamma-\beta)/\alpha},
\end{equation}
in which we chose the inner slope $\gamma$=3/2, outer slope $\beta$=9/2, steepness $\alpha$=1, scale radius $r_0$=1 and total mass of unity (hence $\rho_0$=2/[3$\pi^2$]). A spherical isotropic DF of this system is obtained using the Eddington inversion formula and can be closely approximated by a \texttt{DoublePowerLaw} DF with the following parameters: $J_0$=1.36, $\Gamma$=1.77, B=6.18, $\eta$=1.61, $h_r$=1.37, $g_r$=1.07, $h_z$=$(3-h_r)/2$, $g_z$=$(3-h_z)/2$ (these parameters are found by the  \texttt{example\_doublepowerlaw} program from \Agama).

We draw $N_\text{stars}=10^3$ tracers from this DF and consider four possible scenarios for model fitting. One can treat this model as \textit{self-consistent} (fit only for the mass and the other 6 DF parameters listed above, and determine the potential from the DF iteratively as described in Section~\ref{sec:scm}), or \textit{non-self-consistent} (6 DF parameters and 5 potential parameters are independent from each other; in this case the DF mass is irrelevant and can be fixed to unity). In the non-self-consistent case, the model would need to recover both the tracer density and the mass density (i.e.\ the potential) from the same data, while in the self-consistent case these are identical, so can be constrained better. Independently of this choice, one can use the \textit{joint} position and velocity distribution to constrain the model (i.e.\ scenario A of a known and uniform SF in Figure~\ref{fig:selection_function}), or the \textit{conditional} velocity distribution given the position (scenario C, unknown SF). Naturally, the amount of available information is lower in the latter case, so we expect the constraints on the potential and tracer density to be less precise. In all cases, we assume to know the two sky-plane positions, but not the distance, and all three velocity components without errors.

Figure~\ref{fig:toy_spherical} shows the result of this exercise in all four cases, expressed as the 16/84\% confidence intervals on the density profile (top row), or the relative uncertainty of the density (bottom row). When the SF is assumed to be known, the tracer density can be constrained very well (cyan curve in the right column), as their positions on the sky plane effectively provide the surface density profile, which has a unique deprojection in a spherical system. If the model is self-consistent, the total mass density is known with the same precision (green curve in the left column), but in the opposite case, it still has a fairly small uncertainty at intermediate radii (cyan curve in the middle column), which rises rapidly towards small and especially large radii. On the other hand, when the SF is unknown, the tracer density is very poorly constrained (red curve in the right column), but the mass density is still determined reasonably well at intermediate radii even in the non-self-consistent case (red curve in the middle column), and even better in the self-consistent case (magenta curve in the left column). As our model of the NSC is partially self-consistent (the total mass profile is determined by stellar tracers plus a central SMBH), we could expect an intermediate case between the latter two scenarios. This exercise demonstrates that one can obtain meaningful constraints on the potential even if the tracer density is not known and there are additional mass components besides the stellar tracers. But of course, the actual NSC model is much more complicated than this toy example, so we now describe the other aspects of our NSC fitting procedure.

\subsection{Technical details}  \label{sec:technical_details}

In this section, we describe various approaches for evaluating the model likelihood. We conducted many experiments and convergence tests, ultimately adopting methods and choosing their parameters to strike a balance between computational efficiency and accuracy.

\subsubsection{Computation of projected DF and density}  \label{sec:projected_df}

As already alluded to, with the NSC modelling, we are in scenario C of an unknown SF. Moreover, although the density profile of stars in the Galactic center has been determined in several studies \citep{Do2013a, Schoedel2014a, Schoedel2018, Fritz2016, Feldmeier2017b, GallegoCano2018, GallegoCano2020}, we chose not to use it directly to constrain our models, but only compare the results \textit{a posteriori}. This means that the likelihood of a given star in the model is given by
\begin{equation}  \label{eq:conditional_projected_df}
\mathcal L_i = \mathfrak{f}_\text{proj}(\boldsymbol V_i \,|\, X_i,Y_i) =
\frac{f_\text{proj}(X_i,Y_i,\boldsymbol V_i)}{\Sigma(X_i, Y_i)},
\end{equation}
where $\Sigma$ is the surface density of tracers at the sky-plane coordinates of each star, $\boldsymbol V_i$ are all available velocity components for each star (\vlos and/or PM).

The \texttt{projectedDF} routine in \Agama calculates the multidimensional integrals in $f_\text{proj}$ (Equation~\ref{eq:projected_df_cubature}) very accurately using the \texttt{cubature} library \citep{Johnson2005}, but this is extremely costly ($\gtrsim 10^4$ evaluations of $f$ per star; recall that each evaluation involves the conversion from $\boldsymbol x, \boldsymbol v$ to actions $\boldsymbol J$). Alternatively, one can employ Monte Carlo integration:
\begin{equation}  \label{eq:projected_df_qmc}
f_\text{proj}^\text{(MC)} = \!\!\sum_{k=1}^{N_\text{samples}}\!\! f\big(X,Y,Z^{(k)}, V_X^{(k)}, V_Y^{(k)}, V_Z^{(k)} \big)\; w^{(k)}.
\end{equation}
Here the coordinates and weights $w$ of samples are chosen in such a way as to minimise the variation of $f\big(\boldsymbol{x}^{(k)},\boldsymbol{v}^{(k)}\big)\, w^{(k)}$, in other words, using the importance sampling strategy detailed in Section~2.4 of \citet{Gherghinescu2026} with some additional improvements. In brief, the position $Z$ is sampled from a fiducial 3d density distribution conditioned on the sky-plane coordinates, i.e., $p(Z\, |\, X,Y) = \rho^\text{(fid)}(X,Y,Z) \big/ \Sigma^\text{(fid)}(X,Y)$, where we choose a plausible fiducial profile for $\rho(\boldsymbol x)$ and approximate it with $\mathcal O(10)$ Gaussians, for which conditional sampling can be performed analytically. The missing velocity components (here only PMs, but in principle \vlos can be treated similarly) are drawn from bell-shaped distributions that are both cuspier and more fat-tailed than a Gaussian. These samples are kept fixed throughout the fitting process, minimizing the impact of Poisson noise in the Monte Carlo integration \citep{McMillan2013}. A key improvement with respect to an earlier implementation \citep{Read2021} is that we use quasi-random (low-discrepancy) numbers instead of ``ordinary'' pseudo-random numbers, further improving the accuracy of integration. We verified that with $N_\mathrm{samples}\sim \mathcal O(10^3)$, the difference in the total log-likelihood of the model compared to the use of \texttt{projectedDF} (which is several dozen times more expensive) is well below 1, meaning that the errors in the Quasi-Monte Carlo integration are unimportant for comparing the relative odds of models across parameter space.

The surface density $\Sigma$ at each star's location, which sits in the denominator of Equation~\ref{eq:conditional_projected_df}, can be computed in several ways. First, it is calculated (in a slightly different way) by both \texttt{projectedDF} and \texttt{moments} functions in \Agama, but given that this involves integration over 4 out of 6 phase-space coordinates, this is even more expensive than the marginalisation over missing velocity components. Alternatively, since the self-consistent modelling procedure described in Section~\ref{sec:scm} already delivers the 3d density of stars $\rho(\boldsymbol x)$ that is used to determine the potential, one can simply integrate it along one missing dimension (distance). This is very cheap as it does not involve the conversion between phase-space coordinates and actions. However, recall that the density is represented by a relatively low-order multipole expansion (\texttt{DensitySphericalHarmonic}). 
This is sufficient to determine the potential accurately, but is not sufficient for the surface density; in order to reduce the error in the total log-likelihood well below the acceptable threshold of unity, we would need to use prohibitively large $\ell_\text{max} \ge 40$. We also experimented with first computing the density $\rho$ on a cylindrical grid in $R,z$ and interpolating it using \texttt{DensityAzimuthalHarmonic} to compute projection integrals along $Z$, but again found that the cost is too high for a grid resolution necessary to keep error in $\log\mathcal L$ small.
In the end, we adopted the following approach: since the surface density varies smoothly with $X,Y$, it can be accurately approximated by a 2d interpolator in $\{\ln R, \vartheta\}$, where $R$ is the projected radius and $\vartheta$ is the polar angle in the $X,Y$ plane. With 15 radial and 6 angular points, this interpolation is sufficiently accurate for our purposes, and the cost of computing the density at $\mathcal O(10^2)$ nodes of the interpolation grid using \texttt{moments} is comparable to the cost of Quasi-Monte Carlo integration over missing dimensions for all $\sim 3000$ stars in the catalogue. Both are somewhat sub-dominant to the cost of iterative construction of the self-consistent model. 

\subsubsection{Computation of velocity distributions}  \label{sec:vdf}

\citet{Sormani2022a} and \citet{Binney2023} avoided the need to deal with SFs entirely by considering only the conditional velocity distributions (thus also following scenario C, additionally modified by making the SF distance-dependent in the former paper), but unlike the approach described above, they represented them as products of one-dimensional VDFs in each velocity component:
\begin{equation}  \label{eq:vdf1d}
\mathfrak{f}_\text{proj}\big(\boldsymbol V \,|\, X,Y\big) \approx \prod_{d=1}^3 \mathfrak{f}_\text{1d}^{(d)} \big(V^{(d)} \,|\, X,Y\big) .
\end{equation}
These VDFs were constructed in several dozen spatial regions (bins) as described below, and the likelihood of each star was evaluated using the VDF of the corresponding bin; in other words, the observational catalogue was not binned, but the model was. \citet{Binney2023} used the \texttt{vdf} routine in \Agama to construct smooth spline-interpolated VDFs by deterministic integration. By contrast, in \citet{Sormani2022a} the VDF in each bin was created using essentially the Monte Carlo approach with importance sampling: first, a fiducial model of the entire system described by a certain DF $f^\text{(fid)}(\boldsymbol J)$ was constructed, then a large number ($\gtrsim 10^4$) of equally-weighted samples were drawn from this DF using the \texttt{sampleNdim} routine, and the weight of each sample with phase-space coordinates $\{\boldsymbol x_k, \boldsymbol v_k\}$ was set to $1/f^\text{(fid)}\big(\boldsymbol J(\boldsymbol x_k, \boldsymbol v_k)\big)$. For each subsequent evaluation of a model with a new DF $f^\text{(try)}$, the VDFs in each spatial bin were assembled as weighted sums of samples whose coordinates fall into that bin, the contribution of each sample being given by the ratio of DFs $f^\text{(try)} / f^\text{(fid)}$. By keeping the sample coordinates fixed from the beginning and only changing their weights, one can reduce the impact of Poisson noise on the likelihood evaluation, similarly to the Monte Carlo integration over missing dimensions in $f_\text{proj}^\text{(MC)}$. A smooth VDF was then obtained from these samples using a kernel density estimation approach. \citet{Binney2024} also used \texttt{sampleNdim} from a fiducial DF with subsequent reweighting, but instead of 1d VDFs, they distributed the samples in the velocity space onto a 3d histogram with the cloud-in-cell method (linear interpolation).

The same Monte Carlo sampling approach to VDF construction (only in the \vlos component) was used in the DF-based model of the external galaxy FCC~170 \citep{GalanDeAnta2023}, but the VDFs in each spatial bin were then converted into Gauss--Hermite moments to compare with the observationally-derived VDFs of unresolved stellar populations. In that study, the VDFs were constructed from the reweighted samples and represented by B-splines, using the infrastructure provided by the \texttt{Target(type="LOSVD")} class in \Agama. This approach was originally developed for Schwarzschild modelling, where the samples would come from individual orbits rather than from a DF, and is described in more detail in Section~2.6 and Appendix~A of \citet{Vasiliev2020}.

\begin{figure*}
\includegraphics{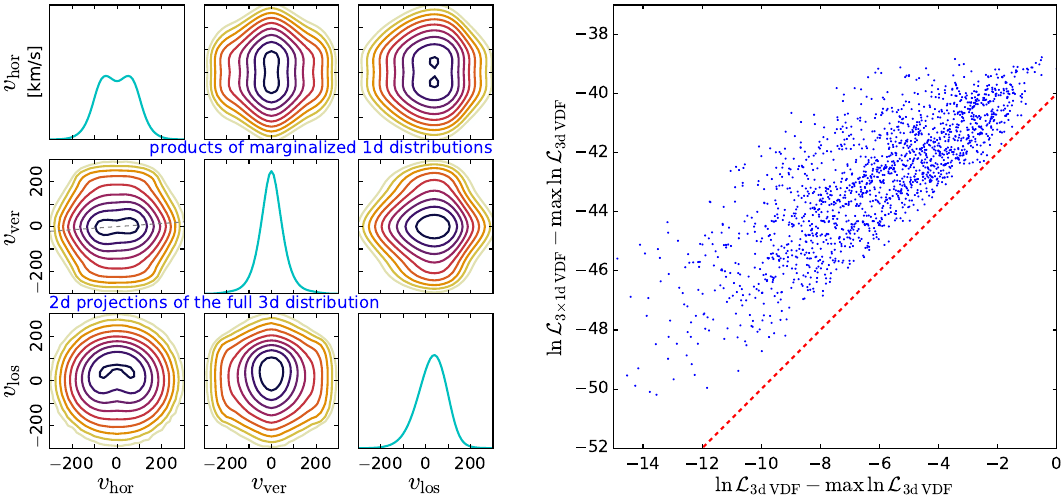}
\caption{\textbf{Left half:} Illustration of the full 3d velocity distribution shown as its 2d projections (bottom left corner) and 1d projections (diagonal panels) at a location $X=3$~pc, $Y=2$~pc from one of the models in the MCMC chain. Details to note are the obviously non-Gaussian shape of marginalised 1d distribution and non-elliptical contours in the 2d distributions, a slight tilt of the $V_\text{hor}$--$V_\text{ver}$ distribution (middle row, left panel), and more prominently, a banana-shaped peak region in the $V_\text{hor}$--\vlos distribution (bottom row, left panel). While the tilt is imprinted in the non-diagonal elements of the velocity dispersion tensor, the banana-shaped peak does not leave a specific signature in it. For comparison, the top right corner shows the approximation of the full velocity distribution (mirrored w.r.t.\ the diagonal from the bottom right corner, i.e., rotated by 90 degrees) as a product of 1d distributions, which lacks these asymmetries. The full 3d velocity distribution is used as the main method of likelihood computation (Section~\ref{sec:projected_df}), while the product of marginalised 1d distributions is an alternative method used in past studies and described in Section~\ref{sec:vdf}.\protect\\
\textbf{Right half:} Comparison of total likelihood values of the entire sample computed with these two methods for a range of models from the MCMC chain. $\ln\mathcal L_\text{3d VDF}$ uses the full 3d velocity distribution for each star (Equation~\ref{eq:conditional_projected_df}), $\ln\mathcal L_\mathrm{3\!\times\!1d\, VDF}$ uses the product of marginalised 1d distributions (Equation~\ref{eq:vdf1d}). Both are shifted by the same constant (the maximum value of the first quantity). The values of log-likelihoods for individual stars computed using both methods are fairly close, with a mean difference of only 0.02 and a scatter of 0.2; however, this nonzero mean offset accumulates over the entire sample to produce an overall shift in log-likelihoods of order a few dozen. There is still a good correlation between the total log-likelihoods computed using both methods, illustrated by the red dotted line with a unit slope, but the scatter of order a few in $\ln\mathcal L_\mathrm{3\times1d}-\ln\mathcal L_\mathrm{3d}$ may affect the relative odds of models in different parts of the parameter space.
}  \label{fig:vdf1d2d}
\end{figure*}

We experimented with the approach of creating binned VDFs from each model, but ultimately abandoned it, because the VDFs change discontinuously between bins and the resulting likelihoods are quite sensitive to the binning scheme. Instead, we opted to interpolate VDFs in the 3d space of projected radius $R$, position angle in the sky plane $\vartheta$, and velocity, creating a 3d cubic spline interpolator for each of the three velocity dimensions. We use the same grid with 15$\times$6 points in $\ln R$ and $\vartheta$ as for the surface density interpolator described earlier. At each grid node (a given point on the sky), we need to compute three one-dimensional VDFs (one in each velocity axis). This can be done by the \texttt{vdf} routine in \Agama, but again the computational costs are excessive (much larger than using the \texttt{moments} routine to compute the surface density). Instead, we again use the Quasi-Monte Carlo approach, at each grid node in $X,Y$ drawing $\sim 10^5$ points in $Z$ and three velocity components with the same importance sampling method as for the missing dimensions in $f_\text{proj}^\text{(MC)}$, and keeping these points fixed throughout the fitting process. For each new model, the VDFs at each grid node are constructed by reweighting the sampling points by the corresponding DF values in the current model, and then interpolating the resulting $\mathfrak{f}_\text{1d}(X, Y, V^{(d)})$ at the location of each star and its corresponding velocity, convolving them with the measurement uncertainties for each star. 

The cost of construction of these VDF interpolators is comparable to the cost of evaluating $f_\text{proj}^\text{(MC)}$ for all stars, as described in the previous section. Unlike the latter, the computational cost is determined by the choice of the grid; the evaluation of interpolated VDFs at each star's location is very fast, so this method may become advantageous if the observational sample size exceeds a few thousand stars.
The likelihood values produced by these two methods are not identical, since the separability of the 3d velocity distribution into a product of 1d VDFs (Equation~\ref{eq:vdf1d}) is only approximate (Figure~\ref{fig:vdf1d2d}), but they are strongly correlated, so that the 1d VDF approach can be reliably used for model fitting. Moreover, if the kinematic data are limited to \vlos only, the two approaches become exactly equivalent (up to numerical errors in computing various marginalisation integrals and interpolation).

\subsubsection{Variable extinction along the line of sight}  \label{sec:3d_extinction}

As mentioned in Section~\ref{sec:fitting_general}, when dealing with the conditional velocity distribution at the given sky position $\mathfrak{f}_\text{proj}(\boldsymbol{V} \,|\,X,Y)$, we made a simplifying assumption that all stars along the line of sight passing through $X,Y$ equally contribute to it. This is not true in general: stars at smaller distances are brighter and thus more likely to be observed (this can be neglected if the intrinsic size of the system is much smaller than the distance to its centre, as is the case for the NSC and even the NSD), and in addition, numerous dust clumps along the same line of sight further suppress the observability of more distant stars. \citet{Sormani2022a} took this effect into account by computing the SF of their catalogue as a function of magnitude, correcting for extinction using a 3d dust map from \cite{Schultheis2014}, and then applying the same SF to the equally-weighted sample drawn from the DF of the model before constructing the 1d VDFs in each spatial field. They report that including the SF did not materially affect the results, but this may be partially attributed to a low spatial resolution of the extinction map.

In the present work, we do not use spatial binning, neither for the observed catalogue nor for the model, so different strategies would be needed if we were to account for the variable extinction along the line of sight. For instance, in Equation~\ref{eq:conditional_projected_df} for $\mathfrak{f}_\text{proj}$ (Section~\ref{sec:projected_df}) one should introduce $Z$-dependent weight factors $S(Z;\, X,Y)$ in the integrals for both the numerator (DF) and denominator (density) along the line of sight. The former is straightforward as the integral in $Z$ at each star's position is computed using the Quasi-Monte Carlo approach: replacing $f$ by the product of $f\,S$ does the job. The denominator, however, is pre-computed on a grid in $X,Y$ and then interpolated at each star's position, so the grid would need to be sufficiently dense to accurately represent the sky-plane variations of the SF $S$. Likewise, the 1d VDFs (Section~\ref{sec:vdf}) are also computed on a grid in $X,Y$ and then interpolated at each star's position and velocity, placing the same demands on the grid spacing. The practical feasibility of these or any alternative approaches remains to be tested.

Instead of modifying the workflow to account for 3d extinction, we address a simpler question: the impact of neglected 3d extinction on dynamical models. \cite{Baes2000} and follow-up studies investigated this in the context of integrated-light kinematic data, and concluded that it may be significant (depending on the orbital structure of the system), but most of the bias occurs in the inferred stellar mass-to-light ratio. In our case, we do not use the observed stellar density profile in any way, so this should not be an issue. The most obvious consequence of 3d extinction is the asymmetry of the $V_\text{hor}$ distribution (Figure~\ref{fig:velocity_distributions} in the Appendix) in some spatial bins, caused by the deficit of stars behind the Galactic centre (which would move with $V_\text{hor}<0$) in the observed sample \citep{Chatzopoulos2015b}. The model VDF is symmetric and thus fits the observed histograms as if they were symmetrised, which, to first order, should not cause much bias. To test this conjecture, we ran the fits on mock data (Section~\ref{sec:mock}) modified to have a larger number of stars in front of the Galactic centre than behind it. Namely, the probability of picking up a star at a distance $Z$ (where negative $Z$ means being in front) was set to $\big[2 - Z / (|Z| + 5\,\text{pc})\big] / 3$, varying between 1 for most nearby stars to 1/3 for most distant ones. Although very crude, this procedure creates a level of asymmetry in $\mathfrak{f}_\text{proj}(\vlos)$ comparable to the actual dataset. The fits to this biased kinematic dataset did not appreciably differ from the fits to the full dataset (all parameters and resulting profiles were compatible at the 0.5$\sigma$ level), so we consider it safe to neglect the 3d extinction.

\subsubsection{Treatment of outliers and foreground contamination}  \label{sec:mixture_model}

\begin{table}
\caption{Velocity distributions of stars in the Galactic bar, which serve as the model for contamination. Each of the three VDFs (longitude, latitude and line-of-sight) is approximated by a sum of two Gaussians $f(v) = \eta_1\,\mathcal N(v \,|\, v_1,\sigma_1) + (1-\eta_1)\,\mathcal N(v \,|\, v_2,\sigma_2)$ with the parameters given below (velocities in \kms).
}  \label{tab:vdf_bar}
\begin{tabular}{lrrrrr}
direction & $v_1$ & $\sigma_1$ & $v_2$ & $\sigma_2$ & $\eta_1$ \\
\hline
$V_\text{lon}$ & $-46$ & 53 & 0 & 169 & 0.28 \\
$V_\text{lat}$ &     4 & 25 & 0 & 131 & 0.21 \\
\vlos          &    42 & 40 & 0 & 146 & 0.04 \\
\hline
\end{tabular}
\end{table}

\begin{figure*}
\includegraphics{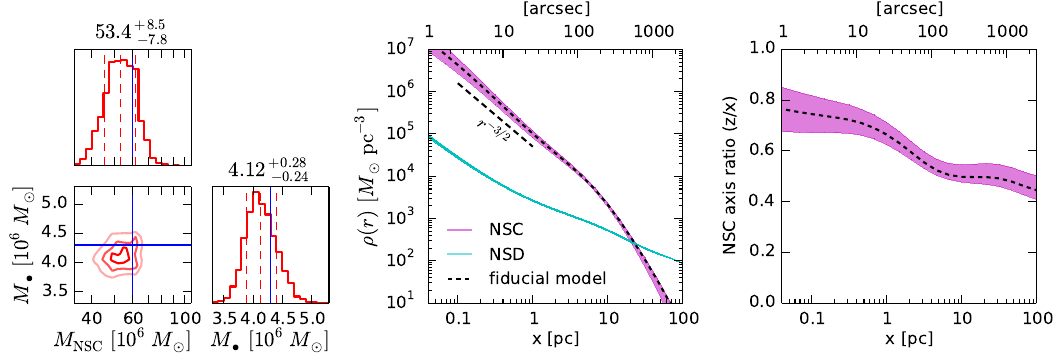}
\caption{Test of modelling machinery on the mock data generated from the fiducial model with parameters given in Table~\ref{tab:fiducial_model}. The left panels show the posterior distributions of the NSC total mass and the SMBH mass. The middle panel shows the 3d density profiles of both NSC and NSD measured along the major axis; the shaded region corresponds to 16th/84th percentiles (note that the NSD DF is kept fixed throughout the fit, and thus its density profile varies very little), and the short-dashed line shows the true profile of the fiducial model. The right panel shows the axis ratio of the NSC in the same way.
}  \label{fig:mock}
\end{figure*}

Although the NSC is the densest region of the Galaxy, it is surrounded by the NSD, and on larger scales, the Galactic bar and the main stellar disc. As mentioned earlier, our model is defined by both NSC and NSD DFs, the parameters of the latter being fixed to the values found in \citet{Sormani2022a}. The surface density of the bar along the line of sight towards the Galactic centre is $\Sigma_\text{bar}\sim 10^4\,M_\odot\,\text{pc}^{-2}$, estimated using the analytic density model from \citet{Sormani2022b}, itself fitted to the made-to-measure $N$-body model of \citet{Portail2017}. This value is only an order of magnitude smaller than the surface density of the NSC+NSD at 10 pc, so the contamination from the Galactic bar is not negligible within the fitted region.

The velocity distribution of the bar component, as shown in Figure~13 of \citet{Sormani2022a}, is nearly constant in the region of interest and can be approximated by a sum of two Gaussians in each velocity axis, with parameters given in Table~\ref{tab:vdf_bar}. The final velocity distribution of the model, from which the likelihood of each star is computed, is a weighted sum of the bar and the NSC+NSD components, weighted by their density at each star's location. When using the projected DF as described in Section~\ref{sec:projected_df}, the likelihood of each star is modified from Equation~\ref{eq:conditional_projected_df} to
\begin{eqnarray}  \label{eq:mixture}
\!\!&& \mathcal L_i = \mathfrak{f}_\text{proj}^\text{mix}\big(\boldsymbol V_i \,|\, X_i,Y_i; \epsilon_{\boldsymbol V,i} \big) \\ \!\!&& = 
 \frac{ f_\text{proj}^\text{(MC)} (X_i, Y_i, \boldsymbol V\!_i; \epsilon_{\boldsymbol V,i}) \;+\; \Sigma_\text{bar} \prod_{d=1}^D \mathfrak{f}_\text{bar}^\text{(d)}\big(V_i^{(d)}; \epsilon_{\boldsymbol V,i}\big) }
{ \Sigma(X_i, Y_i) \;+\; \Sigma_\text{bar}\hspace{13mm} } .  \nonumber
\end{eqnarray}
Here, the second term contains the product of 1d VDFs of the bar for the $D$ velocity components available for the $i$-th star (\vlos and/or two sky-plane components). Both the projected DF in the model $f_\text{proj}^\text{(MC)}$ and the bar VDFs are convolved with the velocity uncertainties of each star $\epsilon_{\boldsymbol V,i}$. When using the separable VDF approximation from Section~\ref{sec:vdf}, the procedure is similar, but $f_\text{proj}^\text{(MC)}$ is replaced by $\Sigma(X,Y)\,\prod_{d=1}^D \mathfrak{f}_\text{1d}^{(d)}\big(V^{(d)} \,|\, X,Y\big)$.

As usual with the mixture modelling approach, one can compute the posterior probability $p_i^{(out)}$ of each star being an outlier (bar contaminant)  as the ratio of the second term in the numerator in Equation~\ref{eq:mixture} to the whole numerator. By summing up these probabilities, we obtain the overall fraction of contaminants in the dataset, which works out to be $\sim$5\% (even though the fraction of stars with $p_i^\text{(out)}\ge 0.5$ is much smaller than that). In principle, the fraction of contaminants (equivalently, their surface density $\Sigma_\text{bar}$) could be made a free parameter optimised during the fit, as in \citet{Feldmeier2025b}, or it could be treated as a second-level parameter in a hierarchical inference procedure (that is, for each choice of the main model parameters, $\Sigma_\text{bar}$ is determined by maximizing the overall likelihood of the model with these parameters and a free bar fraction). We have experimented with the latter option, and the inferred value of $\Sigma_\text{bar}$ turned out to be close to the expected value of $10^4\,M_\odot\,\text{pc}^{-2}$ across a wide range of models, so for simplicity we fixed it at this level.

\subsubsection{Parameter search}  \label{sec:mcmc}

We explore the 12-dimensional parameter space of the model (Table~\ref{tab:fiducial_model}) using the Markov Chain Monte Carlo approach, as implemented in the \textsc{emcee} code \citep{ForemanMackey2013}. We use 30 walkers and run models for $\gtrsim 2000$ steps, visually inspecting the chains for convergence and discarding the initial part of the chain before it stabilises. These runs use a two-level parallelisation strategy, in which different walkers can be considered in parallel by different processes (e.g., 8; here all processes run on the same machine, but in principle they can also be distributed across a cluster using \textsc{MPI}), and each walker uses many (e.g., 16) processor cores via \textsc{OpenMP}. The cost of model evaluation is split roughly 2:1 between the iterative construction of the self-consistent model (Section~\ref{sec:scm}) and the computation of the likelihood using either of the two approaches detailed in Sections~\ref{sec:projected_df} or \ref{sec:vdf}, and takes $\sim$20 seconds on 16 processor cores.

\subsection{Test on mock data}  \label{sec:mock}

To validate our dynamical modelling procedure, we performed a test run on a mock dataset having the same spatial footprint as the observational data described in Section~\ref{sec:data}. We created a fiducial model with parameters given in Table~\ref{tab:fiducial_model}, and for each star in the observational sample, assigned its missing $Z$ coordinate from the conditional distribution $p(Z) = \rho(X,Y,Z) / \Sigma(X,Y)$, and all velocity components that have been actually measured from the conditional distribution $\mathfrak{f}\big(\boldsymbol v \,|\, \boldsymbol x\big)$ composed of both the DF of the dynamical model and the VDF of the bar contamination, mixed in proportion to their surface densities (Equation~\ref{eq:mixture}). Then we perturbed each velocity by a normally distributed noise with the uncertainty taken from the actual observational catalogue.

Figure~\ref{fig:mock} shows the posterior distribution of the NSC and SMBH mass, as well as various derived properties, such as the radial profiles of density, enclosed mass, and axis ratio. Overall, all parameters of the model are recovered within $1\sigma$ confidence intervals, demonstrating the feasibility of dynamical modelling using only the kinematic, but not the density information, in a more realistic situation than the toy models of Section~\ref{sec:toy_examples}.

\section{Observational data}  \label{sec:data}

\begin{figure}
\includegraphics{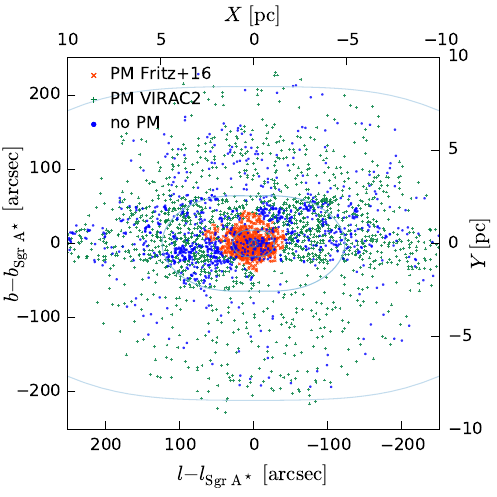}
\caption{Spatial coverage of the observational dataset, which is essentially the same as shown in Figure~1 of \citet{Feldmeier2025b}, but cropped to the central 10 pc. Stars with PM measurements from \citet{Fritz2016} are shown by red crosses, with PM from the VIRAC2 catalogue \citep{Smith2025} -- by green pluses, and with only line-of-sight velocity data -- by blue dots. Blue contours show the combined surface density of the NSC and NSD in the fiducial model, log-spaced by a factor of 2.5 (i.e., one stellar magnitude).
}  \label{fig:data_coverage}
\end{figure}

\begin{figure}
\includegraphics{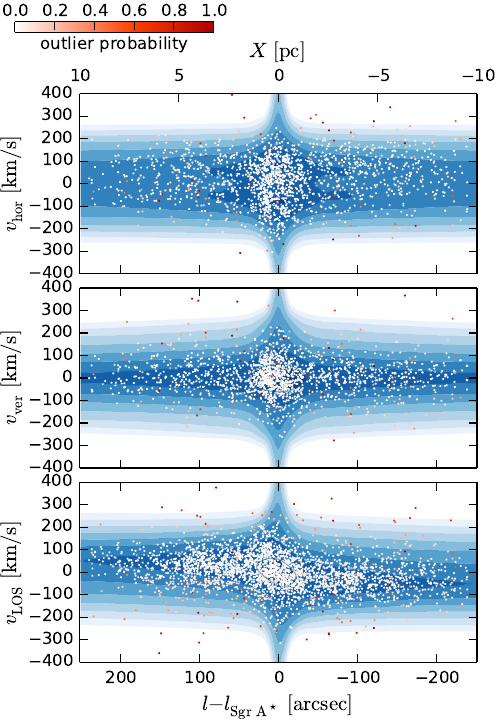}
\caption{Position--velocity diagrams of stars with kinematic measurements after applying quality cuts described in the text. Top panel shows the horizontal velocity $V_\text{hor}\equiv D_0\,\mu_l$, middle panel -- vertical velocity $V_\text{ver}\equiv D_0\,\mu_b$, and bottom panel -- line-of-sight velocity \vlos. The conversion from PM to velocity uses $D_0=8.2$~kpc. Stars are coloured by the probability of being kinematic outliers (i.e. belong to the contaminating population of the bar) computed in the fiducial model. Shaded contours show the distribution function of NSC+NSD in the corresponding velocity component in the fiducial model, log-spaced by a factor of 2.5.
}  \label{fig:data_kinematics}
\end{figure}

\cite{Feldmeier2025b} compiled a catalogue of \vlos and PM measurements from several sources, which we use as a starting point for the present study and apply various additional cuts described below.

The spectroscopic dataset is taken from several sources. \cite{Feldmeier2017a} and \cite{Feldmeier2020} observed spectra of $\sim$700 stars each with KMOS/VLT (\citealt{Sharples2013,Davies2013}, R$\sim$4000) out to 127\arcsec\ ($\sim$5\,pc). Xu et al.\ (in prep.) observed $\sim$900 stars out to 247\arcsec\ ($\sim$9.6\,pc) projected distance with KMOS/VLT. Unlike \cite{Feldmeier2025b}, we added data from \cite{Fritz2016}, who gathered AO-assisted SINFONI/VLT \citep{Eisenhauer2003,Bonnet2004} data with a spectral resolution of R$\sim$1500 and R$\sim$4000. The data for \textgreater\,2\,500 stars  extend to a projected distance to \SgrA of 90\arcsec ($\sim$3.5\,pc). 
Since these data sets overlap, we match stars if they are within 0\farcs5 of each other, and their reported $K_S$-band photometry agrees within 0.6\,mag. If a star has several reported values for \vlos, we match the stars only if their \vlos agree within 30\,\kms, and we compute the mean value for \vlos.

The PM data in the innermost $\sim$3~pc were measured by \cite{Fritz2016} using AO observations from NACO/VLT \citep{Rousset2003,Lenzen2003}, MAD-CAMCAO/VLT \citep{Marchetti2004,Amorim2006} and Hokupa'a/Gemini North \citep{Hodapp1996,Graves1998} with a baseline of $\sim$10~yr. These are complemented by the non-AO catalogue of PMs from the VVV survey (VIRAC2; \citealt{Smith2025}), which extends across the entire region of LOS data, except the most crowded central pc. Some stars are in common between the two catalogues, and their PM generally agree within quoted uncertainties; since the precision of VIRAC2 is much lower than the AO-assisted Fritz et al.'s data (median uncertainty is 0.8\,\masyr and 0.14\,\masyr respectively), we use the former only for stars not in the latter catalogue, rather than combine both measurements for the overlapping subset of stars.

Many stars with \vlos data have no PM measurements and vice versa. It is straightforward to use the union of all datasets for model fitting, marginalizing over the missing velocity components as needed. However, following \cite{Feldmeier2025b}, we chose to use only stars with spectroscopic measurements (and thus \vlos, with or without PM), since this allows us to filter out spectroscopically classified hot young stars, whose kinematics and spatial distribution might differ from the dominant relaxed old population. We further select stars brighter than $K_S=16$ and with the colour index $H-K_S$ between 1.3 and 3.5 (see Figure~2 in their paper for an illustration) to exclude foreground and background stars.
In addition, we apply the following quality filters: \vlos uncertainty less than 15\,\kms, PM uncertainty below 0.3\,\masyr ($\sim$12\,\kms) for the \cite{Fritz2016} data or 1.0\,\masyr ($\sim$40\,\kms) for the VIRAC2 data. Lastly, we restrict the sample to 250\arcsec\ from \SgrA ($\sim$10 pc, the circular area in Figure~1 of \citealt{Feldmeier2025b} mostly covering the NSC, excluding the long and narrow strip along the equatorial plane that extends further into the NSD region). This leaves us with 3\,170 stars having \vlos measurements, of which $\sim$800 have PMs from \cite{Fritz2016} and additionally $\sim$1\,000 have VIRAC2 PMs. 

Figure~\ref{fig:data_coverage} illustrates the spatial coverage of our datasets, and Figure~\ref{fig:data_kinematics} shows the three velocity components for each star plotted against the X coordinate (Galactic longitude). The shaded contours depict the kinematic profiles of the fiducial model, which highlight the key features in the data. The PM distribution is broader in the horizontal direction than in the vertical one, partly due to the vertical flattening of the NSC, and partly due to the overlapping population of the NSD stars, which move up to 100\,\kms in both directions, creating a characteristic profile of $V_\text{hor}$ distribution with two humps and a central depression (see also Figure~\ref{fig:velocity_distributions} in the Appendix for detailed histograms). The \vlos profile shows a clear rotation signature, with receding stars ($\overline{\vlos}>0$) on the left ($X>0$) and vice versa. A small fraction of stars in all three panels are classified as kinematic outliers (bar contamination) in the course of mixture model fitting (Section~\ref{sec:mixture_model}); we stress that we do not filter them out beforehand.

\section{Results and discussion}  \label{sec:results}

\begin{figure*}
\includegraphics{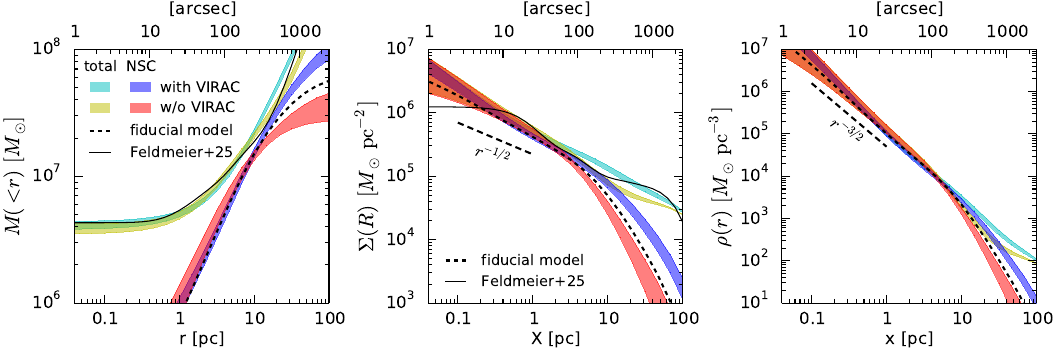}
\caption{Structural properties of the models. Blue and cyan shaded regions show the \V series of models fitted to a combination of \citet{Fritz2016} and VIRAC2 \citep{Smith2025} PM data, red and yellow show the \F series fitted only to the former catalogue. 
The left panel shows the enclosed mass profile as a function of spherical radius (red/blue -- NSC only, cyan/yellow -- also including NSD and SMBH), the middle panel shows the surface density along the projected major axis ($X$), and the right panel shows the 3d density along the major axis ($x$). The shaded region corresponds to the 16/84 percentiles in MCMC. Black dotted lines show the fiducial model with parameters given in Table~\ref{tab:fiducial_model}, and black solid lines show the profiles from \citet{Feldmeier2025b}; see their Figure~7 for comparison with other literature.
}  \label{fig:density_profile}
\end{figure*}

\begin{figure}
\includegraphics{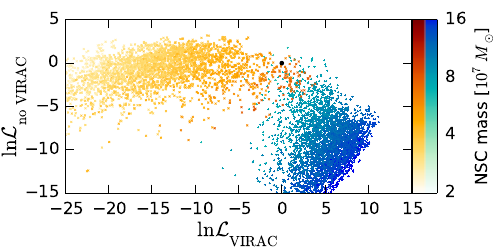}
\caption{Comparison of log-likelihoods of \F (red/orange) and \V series of models. Horizontal/vertical axes show the log-likelihoods of the entire dataset (including VIRAC2 PM) and the kinematic data excluding VIRAC2 PM, which are the objective functions in the \V and \F series, respectively. Values are shifted by a constant, so that the fiducial model (marked by a black dot) is at the origin.
There is only a marginal overlap between the two series of models, and the fiducial model is chosen as a middle ground between them. Points are coloured according to the total mass of the NSC; higher values are in the bottom right corner (i.e., have a higher likelihood with VIRAC2 PMs, but lower likelihood without them).
}  \label{fig:likelihood_comparison}
\end{figure}

\begin{figure*}
\includegraphics{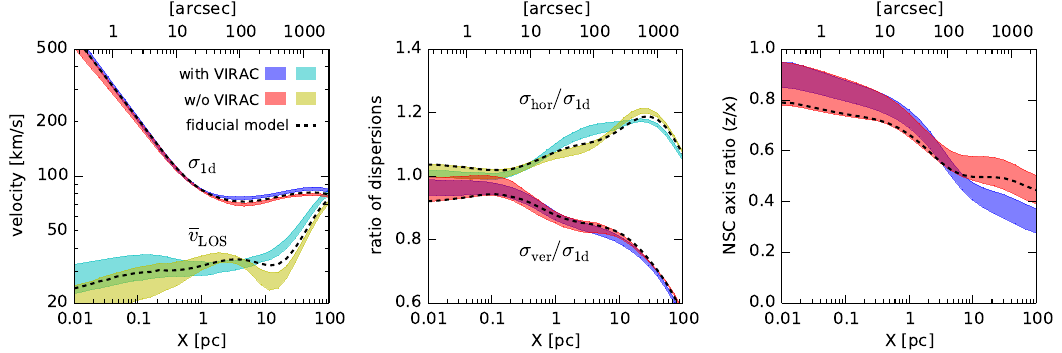}
\caption{Kinematic properties and shape of the \V series (blue and cyan) and \F series (red and yellow) of models, as well as the fiducial model (black dotted lines). Left panel: 1d velocity dispersion (mean of all three components, top curves) and mean \vlos (bottom curves) as functions of radius along the projected major axis ($X$). Middle panel: the ratio of horizontal (top) and vertical (bottom) velocity dispersions to $\sigma_\text{1d}$. The horizontal dispersion at large radii is higher than the other two components, because the velocity distribution is bimodal, with symmetric peaks offset from zero corresponding to stars moving sideways at the near and far sides of the NSD (see e.g. figure~13 in \citealt{Sormani2022a}), while the vertical dispersion is lower due to the flattening of the NSD. Right panel: axis ratio $z/x$.
}  \label{fig:velocity_profiles}
\end{figure*}

We ran model fits for two choices of data: \F series including only the \citet{Fritz2016} PM data (mostly limited to the central 2 pc), or \V series additionally including the VIRAC2 PMs \citep{Smith2025} for stars not in Fritz et al.'s catalogue. In both cases, we also use \vlos measurements for all stars.
It turns out that the constraints on the NSC properties are somewhat different between the two series, although compatible at the $3\sigma$ level.

Figure~\ref{fig:density_profile} shows the 3d density $\rho(r)$ and 2d surface density $\Sigma(R)$ in both series of models, as well as the enclosed mass profile. These profiles match fairly well in the inner region, although the \F series prefer somewhat lower central density slope, closer to $r^{-3/2}$ (and the surface density to follow $\Sigma \propto R^{-1/2}$), while the \V series favour steeper slopes close to $r^{-2}$, which are at odds with photometric constraints on the density profile. For instance, \citet{GallegoCano2018} and \citet{Schoedel2018} found the power-law exponent of the surface density within 1~pc to be in the range 0.3--0.5, whereas \citet{GallegoCano2020} fitted the 3d density by a double-power-law profile with an inner slope of 1.3--1.45, depending on the selection. 

More significant is the difference in the behaviour of \F and \V series in the outskirts (beyond $\sim$10 pc, which is the extent of our kinematic data). In the former case, the break in the density profile occurs at smaller radii ($\sim$5~pc), although it still drops rather gradually, so that roughly half of the mass lies beyond 10~pc. In the \V series, the cluster extends further out, and the density drops even less steeply, and the total mass of the NSC is $\gtrsim$ $10^8\,M_\odot$. We remind that in the present study, we keep the DF of the NSD fixed, and if the kinematic data require a larger mass in the outskirts of the model, it has to be assigned to the NSC. This interpretation is supported by the fact that the NSC becomes more flattened in the outer region, with axis ratio $q \lesssim 0.5$. In future work, it might be interesting to rerun the fits with newer kinematic data beyond 10~pc and optimise the parameters of both NSC and NSD.

Figure~\ref{fig:likelihood_comparison} compares the likelihoods of both model series when including or excluding VIRAC2 PMs. The \F series (red/yellow) have a higher likelihood when excluding VIRAC2 (vertical axis), but lower when including them (horizontal axis); we remind that these models were optimised against the former dataset. Conversely, the \V series have a higher likelihood when using the entire kinematic data, including VIRAC2, since they were optimised against this dataset, but a lower likelihood when excluding VIRAC2. There is a marginal overlap between the two series of models, and the fiducial model was picked from that region of parameter space (black dot). Among all parameters, the total mass of the NSC has the clearest trend with the likelihood of both datasets: VIRAC2 PM and hence the \V series prefer higher masses, and the remaining kinematic data are consistent with lower masses.

Figure~\ref{fig:velocity_profiles} illustrates the kinematic properties of models, shown as the mean \vlos and the velocity dispersion profiles along the major axis. All three components of velocity dispersion exhibit a Keplerian rise ($\sigma \propto r^{-1/2}$) at radii $\lesssim 1$~pc. Outside this radius, there is a notable anisotropy between velocity dispersions in the horizontal direction (largest), vertical (smallest), and line-of-sight (intermediate; this dimension also has a nonzero mean velocity that reaches a few tens \kms in the NSD region). Although the differences between the two series of models are only marginal, they apparently drive them apart in the parameter space. We remind that the VIRAC2 PMs are used only in the region 2--10 pc, where the curves start to diverge.
The NSC DF in both series of models has a near-maximal rotation amplitude $\varkappa$, producing a mean $\vlos\simeq30$\,\kms within the central 10~pc (bottom curves in the left panel). This agrees well with the literature results (e.g., figure~10 in \citealt{Feldmeier2025b}).
We also note that inside the central parsec, the models are close to isotropic (central panel; the plot shows the projected velocity dispersions, but the intrinsic ones are also very close to each other). Although our adopted \texttt{DoublePowerLaw} DF can produce anisotropic models by adjusting the mixing coefficients $h_{r,z}$ in Equation~\ref{eq:df_nsc}, the ratio of velocity dispersions is fairly moderate even for rather extreme values of these coefficients, particularly when the inner slope $\Gamma$ is close to zero (equivalently, when the density slope is close to 3/2). This is likely a limitation of this DF family, but to test this hypothesis, one would need to repeat the fitting experiment with an alternative choice of DF. 

As a consequence of the weak sensitivity of anisotropy to model parameters, the axis ratio of the NSC also stays close to 1, although in the \F series, one can find models with the inner axis ratio $z/x$ anywhere between 0.6 and 1 with very little variation in the likelihood (top right panel in Figure~\ref{fig:corner_plot} in the Appendix). For the fiducial model, we picked up parameters producing an inner axis ratio around 0.8 (Figure~\ref{fig:velocity_profiles}, right panel), which is only slightly higher than the photometrically determined value $\sim$0.7 \citep{Schoedel2014a, Chatzopoulos2015a, GallegoCano2020}.

The mass of the SMBH in both series of models is similar: \mbh=$(3.9\pm0.4) \times 10^6\,M_\odot$ in \F and $(4.1\pm 0.3)\times 10^6\,M_\odot$ in \V, which agrees within uncertainties with the results of \citet{Feldmeier2025b} ($[4.35\pm 0.24]\times 10^6\,M_\odot$), as well as with the ``true'' value determined from the multi-year monitoring of individual stars very close to the SMBH ($4.3\times 10^6\,M_\odot$, \citealt{Gravity2022}). This is reassuring, but by no means a trivial result. In fact, throughout most of this project, the best-fit value of \mbh was systematically lower by as much as a third, causing endless consternation and prompting further tests and alterations of the analysis pipeline. Only after incorporating a more realistic contamination model (Section~\ref{sec:mixture_model}) with the velocity distribution taken from a bar model with a correct normalisation, the best-fit SMBH mass drifted close to the true value. In earlier versions of the modelling pipeline, we used a spatially-independent fraction of contaminants, while in the current version, the absolute normalisation of the contaminant population is spatially constant, but the relative fraction is lower in the denser central regions. As a consequence, stars in the tails of the velocity distribution are less likely to be attributed to the bar contamination if they are closer to the SMBH, and thus require a larger \mbh to accommodate their high velocities. This high sensitivity of \mbh to seemingly minor details of the modelling procedure is a cautionary tale for SMBH mass measurements in external galaxies, in which the quality of observational data is nowhere near that of the Milky Way NSC, but one does not have the luxury of knowing the true answer.

\begin{table}
\caption{Total enclosed mass at several radii (in units of $10^6\,M_\odot$) in the \F and \V series, compared to the free-\mbh model from \cite{Feldmeier2025b}, which uses kinematic data extending up to 33~pc.}  \label{tab:mass_profile}
\begin{tabular}{lccc}
model     & $M(<5\text{ pc})$ & $M(<10\text{ pc})$ & $M(<20\text{ pc})$ \\
\hline
\F series & $12.1 \pm 0.7$    & $21.0 \pm 1.1$     & $38 \pm 3$ \\
\V series & $11.2 \pm 0.6$    & $22.2 \pm 1.1$     & $50 \pm 3$ \\
FK+25     & $11.8 \pm 0.3$    & $18.7 \pm 0.5$     & $37 \pm 1$ \\
\hline
\end{tabular}
\end{table}

The constraints on the extended mass distribution around the SMBH are also similar between both series up to $\sim 10$~pc, and are compatible with earlier studies. For instance, \citet{Magorrian2019} determined the stellar mass within 4~pc to be $(6.5\pm 0.6)\times 10^6\,M_\odot$, whereas in the \F series it is $(6.3\pm 0.7)\times 10^6\,M_\odot$ and in the \V series is slightly lower at $(5.3\pm 0.5)\times 10^6\,M_\odot$. The total enclosed mass at several values of radii is listed in Table~\ref{tab:mass_profile}; the \F series gives very similar results to \cite{Feldmeier2025b}, while the \V series starts to diverge beyond 10~pc.

Our model for the NSC and NSD has only a single stellar population, and we do not use any chemical or age information to disentangle the contributions of stars with different metallicities. Using two-population Jeans axisymmetric models, \citet{Feldmeier2025b} found that over 90\% of mass belongs to the metal-rich component, and moreover there is little difference in the inferred mass distribution between one- and two-component models. Adding the chemical dimension to DF-based models is, in principle, straightforward, either as a combination of two or more DFs \citep[e.g.,][]{Bovy2013,ArroyoPolonio2025} or using the ``extended DF'' formalism with the structural parameters of the DF depending on chemistry \citep{Sanders2015a,Das2016,Binney2024}. We also ignored the contribution of dark matter to the total density, as \citet{Feldmeier2025b} found it to be negligible in the central region; again, incorporating it into DF-based models is straightforward and has been done previously \citep{Piffl2015,Binney2015,Cole2017,GalanDeAnta2023,Binney2023}.

Given that the \F and \V series prefer somewhat different regions in the parameter space, we picked up the fiducial model roughly at the intersection of these regions, taking the liberty to choose parameters that better match the observational constraints on the stellar density profile (inner slope and flattening), given that the models do not use this information directly. This fiducial model is marked by dashed lines in the above plots, and was used in Section~\ref{sec:mock} to test the accuracy of recovery of model parameters. We consider it to be an adequate representation of the NSC given the observational data used in this study, but in future work, we plan to incorporate additional kinematic data beyond 10 pc and allow the parameters of the NSD to be optimised in the fit, rather than keeping them fixed to those determined in \citet{Sormani2022a} as we did here. We make the fiducial model publicly available as a script \texttt{example\_mw\_nsc.py} included in the \Agama repository. It can be used to visualise the velocity distributions at different spatial locations, create representative samples of phase-space points drawn from the DF, integrate orbits, etc.

\section{Summary}  \label{sec:summary}

In this work, we developed a DF-based approach for constructing dynamical models of stellar systems constrained by discrete-kinematics data, accounting for possible foreground contamination and imposing dynamical self-consistency for the stellar distribution. We tailored our modelling workflow to the specific case of the Milky Way nuclear star cluster, but the general approach can be applied to any resolved stellar system. We explored various solutions to technical problems that made our implementation computationally feasible while retaining sufficient accuracy of the likelihood (Section~\ref{sec:technical_details}).
We demonstrated that it is possible to obtain meaningful constraints on the mass density profile, even if we do not use any information about the spatial distribution of tracers, only their velocity measurements (Sections~\ref{sec:toy_examples}, \ref{sec:mock}). Although in reality we are rarely in a situation when the density profile of tracers is completely unknown, a strongly spatially varying dust extinction can make its determination challenging, whereas using a dynamical model, one can predict the spatial distribution of tracers from kinematics and then in turn use it to constrain the dust distribution \citep[e.g., section 5.7 in][]{Binney2023}.

We then applied our method to the actual observational data ($\sim$3\,200 stars within 10~pc from the Galactic centre, of which $\sim$1\,800 have PM measurements and the rest have only \vlos data; Section~\ref{sec:data}). In this study, we fix the parameters of the NSD to those found by \citet{Sormani2022a}, and only vary the parameters of the NSC DF and the mass of the central SMBH.
As discussed in Section~\ref{sec:results}, we found noticeable differences between two kinematic datasets (\F series using only the PM data from \citet{Fritz2016} in the central $\sim$2~pc, or \V series adding PMs from the VIRAC2 catalogue, \citealt{Smith2025}). Most apparently, the \V series demand a higher total mass of the NSC ($\gtrsim 10^8\,M_\odot$), most of which resides well outside the central 10~pc region. We conjecture that this may result from an underestimation of the NSD contribution in this region, and plan to rerun our models incorporating additional observational data beyond 10~pc and varying the NSD parameters as well. At the same time, the NSC properties in the central 10 pc largely agree between both \F and \V series, and are compatible with prior literature. We thus believe that our fiducial model, chosen to be marginally compatible with both series, is a good representation of the NSC structure in this region.

We also verified that our models can recover the mass of the central SMBH: $\mbh \simeq (4\pm0.4) \times 10^6\,M_\odot$, which is compatible with the much more precise constraints provided by \citet{Gravity2022}. It should not be taken as granted, though: historically, many dynamical studies of the NSC \citep[e.g.,][]{Feldmeier2014,Feldmeier2017b,Fritz2016} yielded lower SMBH masses than obtained from fitting of individual orbits of S-stars (\citealt{Ghez2008,Genzel2010}, and more recently \citealt{Gravity2018,Gravity2019,Gravity2020,Gravity2022}). Only in some recent equilibrium models of the NSC \citep{Chatzopoulos2015a,Magorrian2019,Feldmeier2025b} the SMBH mass approached $4\times10^6\,M_\odot$, and in our models, we were able to get a reasonable value only after adding a nontrivial treatment of outliers (contamination from the Galactic bar). This highlights the difficulty of measuring \mbh from single-epoch stellar kinematics even when using thousands of stars within the influence radius, far more than typically available for constraining IMBHs in globular clusters \citep[e.g.,][]{DellaCroce2024,Haberle2024,BanaresHernandez2025}, not to mention external galaxies with only integrated-light kinematics.

Our modelling approach also has a number of limitations, some of which arise from the specific choice of the DF for the NSC (e.g., the difficulty of creating strongly anisotropic velocity distributions when the density profile near the SMBH is close to $r^{-3/2}$), and other are more general (e.g., the restriction to axisymmetry in the current implementation).
Despite these, we believe that DF-based self-consistent models are well-suited for determining mass distribution from discrete kinematics, complementing other modelling approaches, and can be easily generalised to include multiple stellar populations and additional (e.g., chemical) information.

\begin{acknowledgements}
We thank the referee for valuable comments and suggestions.
EV acknowledges support from an STFC Ernest Rutherford fellowship (ST/X0040066/1).
AFK acknowledges funding from the Austrian Science Fund (FWF) [grant DOI 10.55776/ESP542].
MCS acknowledges financial support from the European Research Council under the ERC Starting Grant ``GalFlow'' (grant 101116226) and from Fondazione Cariplo under the grant ERC attrattivit\'a no.\ 2023-3014.
This research was conducted organically, i.e., without AI assistance, neither in coding nor in manuscript preparation.
\end{acknowledgements}


\appendix



\begin{figure*}[b]
\includegraphics{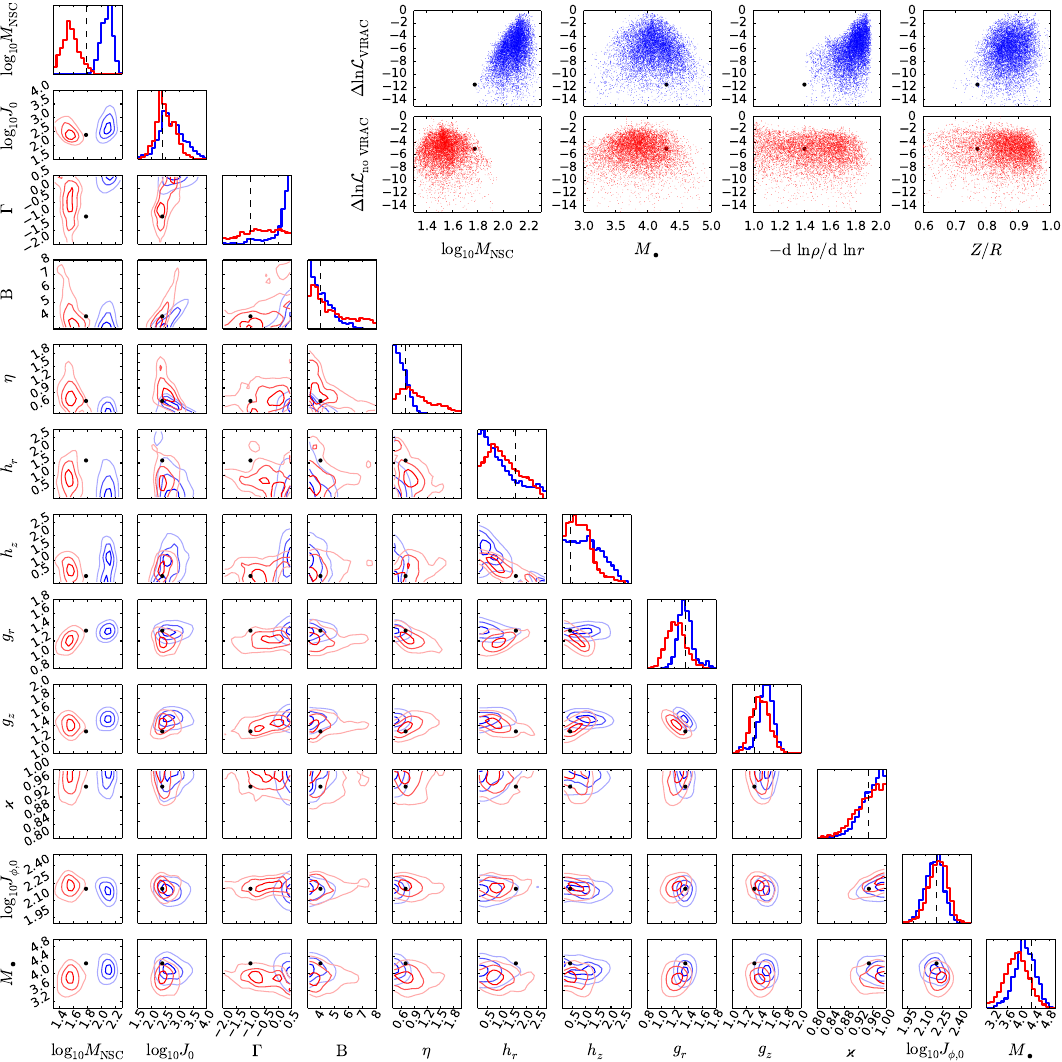}
\caption{Corner plot of all 12 model parameters (see Table~\ref{tab:fiducial_model} for their meaning) in the \V series (blue) and \F series (red). Additional panels in the top row show the offsets in log-likelihoods from the maximum value for all models from the MCMC chain as functions of several parameters: total NSC mass, SMBH mass, density slope and axis ratio (the latter two evaluated between $r=0.01$ and 0.1~pc). For instance, although the axis ratio in both model series mostly lies between 0.8 and 1 (Figure~\ref{fig:velocity_profiles}, right panel), the maximum likelihood of models in the \F series is largely independent of the axis ratio. The fiducial model parameters are marked by a black dot, or by a vertical dashed line in the 1d histograms.
}  \label{fig:corner_plot}
\end{figure*}

\begin{figure*}
\includegraphics{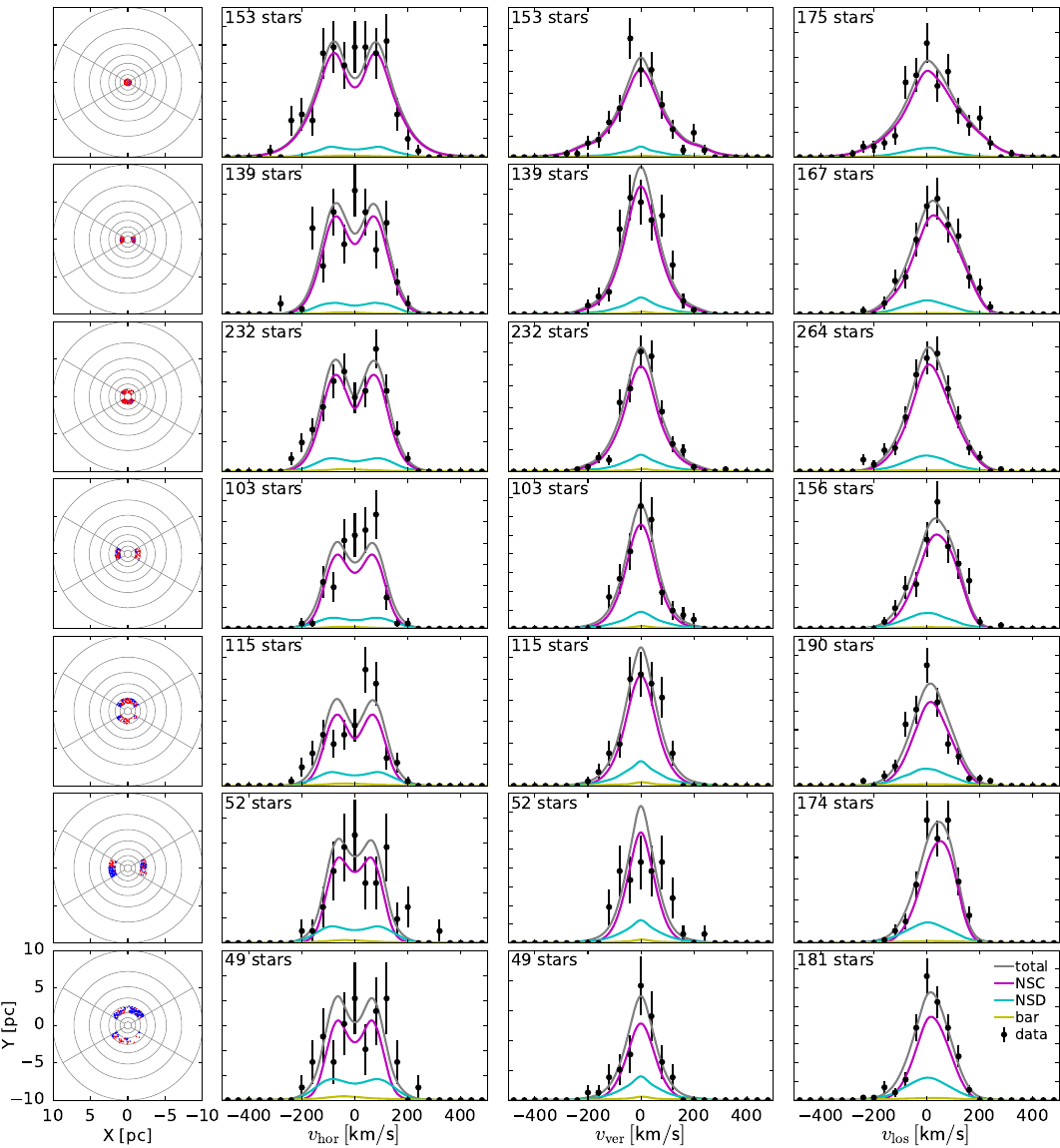}
\caption{Velocity distributions in the fiducial model (smooth curves) compared to the observed velocity histograms. Each row corresponds to one spatial region, as shown in the left column (red/blue points are stars with/without PM). In the remaining columns, the observed velocity distributions of stars in this region ($V_\text{hor}$, $V_\text{ver}$, \vlos) are shown by histograms with Poisson uncertainties (black), and the VDFs in the model (computed at the location of each star and averaged over all stars) are shown by curves of different colour: NSC (magenta), NSD (cyan), bar (yellow), and total (gray). The \vlos distribution is antisymmetric w.r.t. the change of sign of $X$, so we flip the signs of $X$ and \vlos for stars with $X<0$ to show both left and right sides of the NSC on the same histogram. The spatial and velocity binning is used only for visualisation purposes; the models are fitted to individual stellar velocities.
}  \label{fig:velocity_distributions}
\end{figure*}

\begin{figure*}
\addtocounter{figure}{-1}
\includegraphics{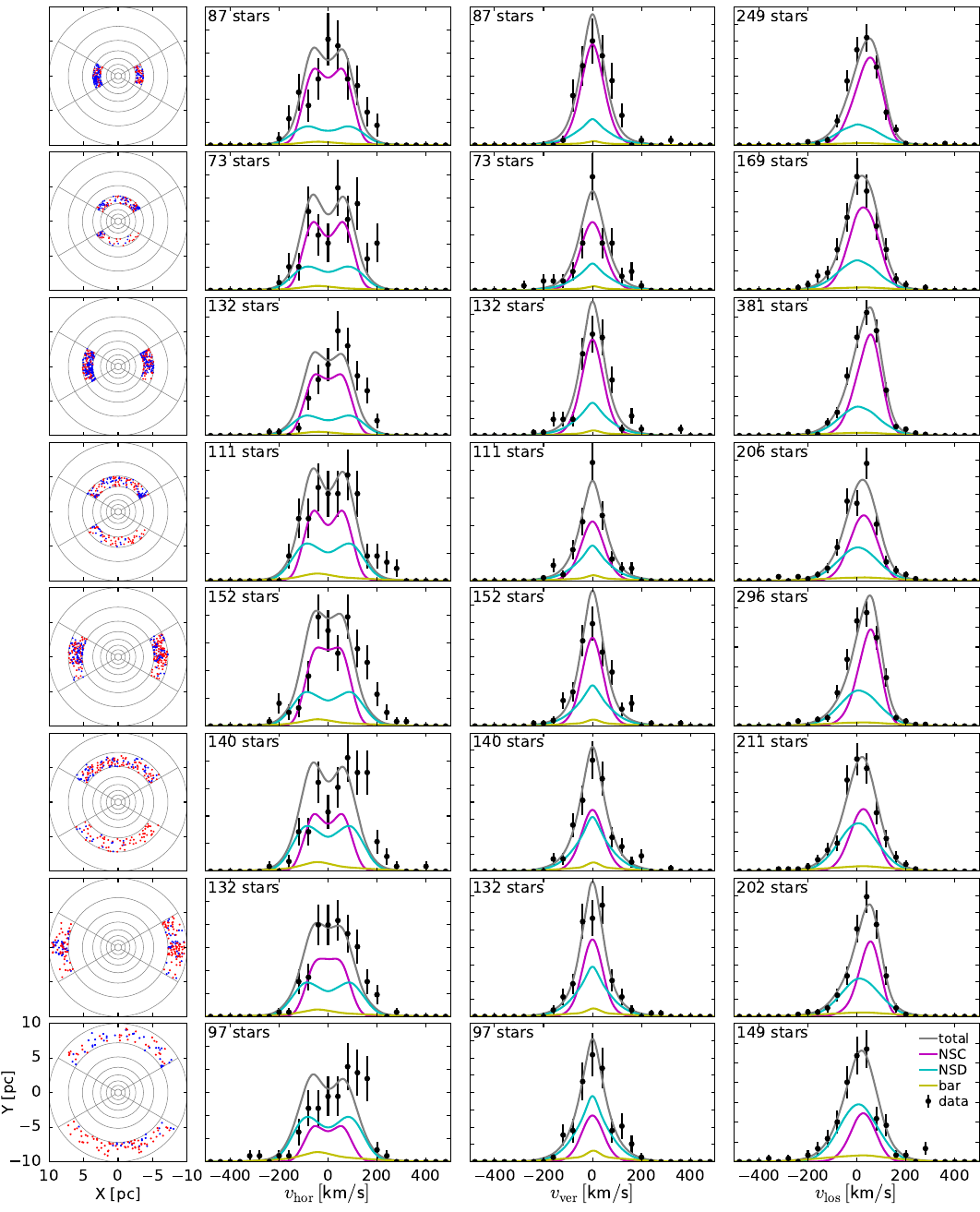}
\caption{Velocity distributions (continued)}
\end{figure*}

\end{document}